\documentclass[12pt]{article}

\usepackage{newtxtext,newtxmath}

\usepackage{graphicx}
\usepackage{here}
\usepackage{url}
\usepackage{array}
\usepackage{makecell}

\usepackage[letterpaper,margin=1in]{geometry}

\renewenvironment{abstract}
	{\quotation}
	{\endquotation}

\date{}

\makeatletter
\renewcommand{\fnum@figure}{\textbf{Figure \thefigure}}
\renewcommand{\fnum@table}{\textbf{Table \thetable}}
\makeatother

\usepackage{scicite}

\usepackage{url}

\def\scititle{
	Integrating Domain-Specialized Language Models with AI Measurement Tools for Deterministic Atomic-Resolution Experimentation
}
\title{\bfseries \boldmath \scititle}

  \author{
      Zhuo~Diao$^{1\ast}$,
      Kouma~Matsumoto$^{1}$,
      Linfeng~Hou$^{1}$,
      Masahiro~Ohara$^{1}$,
      Hayato~Yamashita$^{2}$\and
      Masayuki~Abe$^{1\ast}$\and
      \small$^{1}$Graduate School of Engineering Science, The University of Osaka, Osaka 560-0043, Japan.\and
      \small$^{2}$Graduate School of Integrated Science and Technology, Shizuoka University, Shizuoka, 432-8561, Japan.\and
      \small$^\ast$Corresponding authors. Email: diao.zhuo.es@osaka-u.ac.jp; abe.masayuki.es@osaka-u.ac.jp
  }

\begin{document} 

\maketitle

\begin{abstract} \bfseries \boldmath
Self-driving laboratories based on large language models promise to transform scientific discovery through general experimental automation. However, realizing this vision on precision platforms remains challenging, requiring deterministic execution and effective domain adaptation under strict physical constraints. We address these requirements through a framework that specializes in small language models for autonomous control of scanning probe microscopy, coordinating task-specific models with AI-driven measurement tools. We demonstrate real-time, atomic-resolution SPM experiments at room temperature, achieving instruction-level control and multi-step experimental planning. Fine-tuning reduces perplexity from 1.44 to 1.20 and improves reliability, with the adapted model reaching 99.3\% and 95.2\% command accuracy, outperforming OpenAI o4-mini on domain-specific tasks. 
This architecture achieves lower computational cost while maintaining deterministic execution and enabling deployment on consumer-grade hardware.
This work bridges probabilistic language models with deterministic experimental control through a modular, domain-specialized architecture, providing a generalizable pathway toward scalable and trustworthy self-driving laboratories across diverse scientific platforms.

\end{abstract}

\noindent
Scientific discovery has traditionally relied on human intuition, manual operation, and accumulated experimental expertise. 
As experimental platforms become increasingly complex and data-intensive, this reliance on expert operators constrains throughput, reproducibility, and the overall pace of discovery. 
These challenges have driven the development of self-driving laboratories (SDLs), in which experiments are autonomously planned, executed, and analyzed with minimal human intervention~\cite{Abolhasani:2023aa, sdl1, sdl2}.
However, for a broad class of precision scientific discovery—particularly those operating at the nanoscale—achieving full autonomy remains an open challenge.

Electron microscopes and scanning probe microscopes (SPMs) are representative examples of such systems, where automation is especially demanding~\cite{Kalinin:2021aa, Liu:2022aa}. 
Their operation relies on highly labor-intensive workflows, in which successful measurements are difficult to reproduce using rigid, predefined protocols. 
Instead, experimental expertise is developed through prolonged training, during which operators learn to interpret ambiguous measurement states and devise appropriate corrective strategies~\cite{RevModPhys.75.949}.
These challenges are further exacerbated in room-temperature SPM experiments. Thermal drift degrades positioning accuracy over time, while the probe--sample interaction energies required for atomic resolution can destabilize the tip state, reducing reproducibility~\cite{Sugimoto:2008aa, Lahiri:2011aa}. Consequently, enabling SDLs on nanoscale platforms requires not only automation of experimental procedures, but also the ability to capture and reliably deploy tacit, instrument-specific knowledge derived from human expertise.

Considerable progress has been made in automating workflows within specific SPM experiments.
Bayesian optimization frameworks accelerate parameter search and measurement~\cite{Liu:2023aa, D4DD00277F, Harris:2025aa}, AI-based pattern recognition enables real-time image feedback~\cite{Diao:2025aa, Sung:2025aa}, and dedicated algorithms address thermal drift~\cite{Diao:2023aa, Deveci:2025aa} and tip conditioning~\cite{Diao_2023, Krull:2020aa}.
While each of these approaches addresses a well-defined experimental challenge with high precision, they typically encode task-specific logic tailored to a single function. 
As a result, coordinating multiple corrective actions, responding to unforeseen experimental states, or interpreting high-level scientific instructions still requires explicit human intervention. This limitation highlights a critical gap between isolated automation tools and the realization of a truly autonomous laboratory agent.

LLMs exhibit strong generalization and decision-making capabilities, making them promising tools for scientific automation. 
Through context engineering at inference time, their effective knowledge boundaries can be extended~\cite{M.-Bran:2024aa, Liu:2025aa}, enabling adaptation to new domains and interpretation of high-level experimental intent. 
Recent approaches, including tool-calling frameworks, retrieval-augmented generation (RAG), and structured skill injection, allow LLMs to plan and execute scientific workflows~\cite{Prince:2024aa}, translating abstract instructions into instrument-level operations~\cite{Boiko:2023aa, Xie:2025aa}. 
In advanced instrumentation, early demonstrations have shown that LLMs can generate control code for SPM systems~\cite{Liu_2024} and guide experiments via prompt-based interaction in both ambient and ultrahigh-vacuum environments~\cite{Mandal:2025aa, gptspm}. 
However, these efforts primarily address whether LLMs can be applied to instrument control, rather than whether they can operate reliably under strict physical constraints. 
In precision systems such as atomic-resolution SPM, control requires determinism, bounded parameter spaces, and real-time responsiveness, where even minor deviations can lead to irreversible experimental failure. 
Existing LLM-agent frameworks, which rely on prompt-based reasoning and cloud-hosted general-purpose models, do not provide mechanisms to guarantee deterministic execution or domain-specialized reproducibility, and their dependence on remote inference introduces latency incompatible with real-time control.

These limitations are not incidental but arise from the underlying design paradigm of context-engineered LLM-agent systems.
Because outputs are generated through probabilistic decoding, identical inputs can yield different command sequences, and hallucinated parameters may occur even when correct rules are present~\cite{xu2025}. 
Such variability is unacceptable in nanometer-scale SPM, particularly in advanced tasks such as atomic manipulation~\cite{Chen:2022aa, Okuyama:2025aa} and probe-assisted reactions~\cite{Su:2024aa, Zhu:2026aa}.  
The integration of multimodal signals further exacerbates this issue when mediated through text-based representations~\cite{Miret:2025aa, Alampara:2025aa}. 
In addition, reliance on large contextual inputs leads to significant computational overhead, limiting efficiency in high-frequency control loops~\cite{shen2024}. 
This challenge is amplified in cloud-based deployments, where latency, data privacy concerns, and per-token costs hinder practical deployment, while large-scale inference incurs non-negligible energy consumption~\cite{edgeai, 10.1145/3630106.3658542}. 
Taken together, these constraints reveal a fundamental mismatch between the probabilistic, prompt-driven nature of existing LLM-agent frameworks and the deterministic, resource-constrained requirements of precision scientific instrumentation. 
Therefore, enabling reliable deployment in such settings requires a shift from inference-time context engineering to architectures that enforce determinism, bounded action spaces, and efficient domain specialization by design.

Here, we propose an LLM-driven automation framework that targets the stringent reliability requirements of scientific instrumentation—requirements that existing prompt-engineering and context-augmentation approaches are structurally unable to satisfy. 
Rather than extending general-purpose LLMs with domain context at inference time, the framework adopts a fine-tuning strategy to specialize small language models (SLMs) and coordinate multiple models with distinct functional roles, enabling the system to operate on a consumer-grade GPU while orchestrating real-time, room-temperature atomic-resolution SPM experiments.
Its capabilities include scheduling instrument-specialized operation interfaces, integrating task-specific AI modules, and issuing low-level instrument commands, as well as formulating and executing experimental plans based on implicit operational knowledge.
Evaluation results show that the fine-tuned models achieve lower perplexity while producing precise and highly reproducible outputs, significantly reducing the risk of hallucination in scientific experiments.
Notably, the system demonstrates superior performance over cloud-based models in awareness of instrument constraints and nanoscale experimental planning under room-temperature conditions.
This LLM-based framework outlines a reliable and generalizable pathway toward self-driving laboratories that bridge the gap between high-level scientific intent and real-time instrument-level execution on precision scientific platforms.
Although demonstrated using SPM as a testbed, the underlying architecture is instrument-agnostic and readily extensible to other scientific platforms.

\section*{Results}

\subsection*{Self-driving experiments via fine-tuned SLM}

\begin{figure}[h]
\centering
\includegraphics[width=0.9\textwidth]{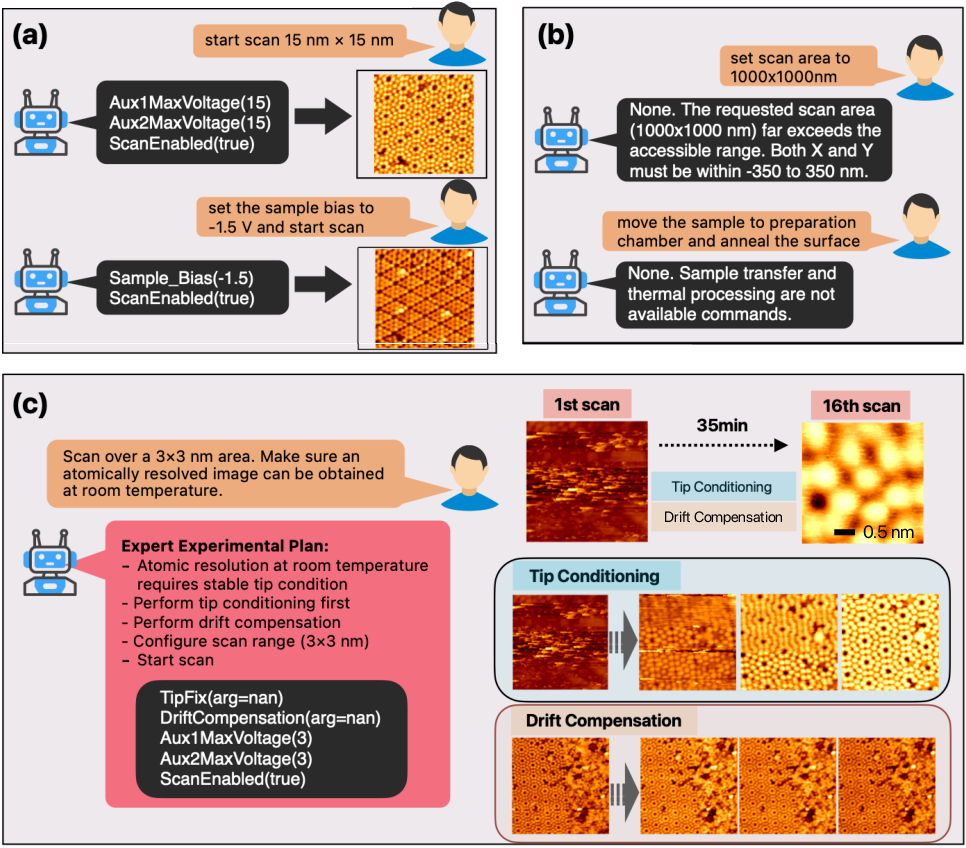}
\caption{
A two-stage autonomy framework enabled by a fine-tuned, domain-specialized small language model (SLM), demonstrated in real-time scanning probe microscopy (SPM) experiments.
The framework visualizes the interaction among user instructions, SLM-generated outputs, and corresponding experimental results. More detailed system outputs and experimental traces are provided in Fig. S2 and Fig. S3 in Supplementary Information.
(a) Stage i: Direct execution of user-issued control commands generated by the SLM.
(b) Rejection of invalid or out-of-specification instructions through constraint-aware validation.
(c) Stage ii: Autonomous formulation and execution of multi-step experimental plans based on high-level user intent.
 }
\label{fig:exp_demo}
\end{figure}

Our fine-tuned SLMs are capable of providing real-time assistance during the experimental process to achieve stable and autonomous operation of SPM experiments in place of human operator actions. 
We deployed the framework in a real-time experimental setting using scanning tunneling microscopy (STM), an implementation of SPMs, to achieve atomic-resolution imaging of a Si(111)-(7$\times$7) surface at room temperature.
Achieving atomic-resolution imaging at room temperature is inherently challenging due to thermal drift and reduced tip stability, and typically requires extensive experimental
expertise and manual operation.
Conventional SPM systems lack built-in mechanisms to directly mitigate these issues and instead rely on expertise accumulated through extensive trial-and-error.
In the proposed SLM-integrated SPM system, the model can orchestrate AI-driven compensation and conditioning modules when required to address these issues autonomously.
Fig.~\ref{fig:exp_demo} shows the results of executing user instructions.
Depending on the model's decision-making capability, we categorize the automation levels of the SLM into two stages:

\begin{itemize}
\item Stage \rm{i}: instruction-driven control with constraint enforcement \\

As shown in Fig.~\ref{fig:exp_demo}(a), the SLM converts natural language instructions into structured SPM commands and returns the measurement results.
This allows the SPM system to be controlled through text input rather than through a standard PC interface.
The complete user interface outputs during the experiments are shown in Fig. S2 and Fig. S3 in the Supplementary Information.
Such capability allows the SLM to actively intervene in experimental operations and can serve as a basis for fully automated experimental systems.
In addition, it unifies instrument operation at the language level and significantly reduces the learning cost across different SPM systems.
The SLM is also designed to enforce constraints imposed by instrument specifications.
As shown in Fig.~\ref{fig:exp_demo}(b), when encountering invalid instructions, such as unreasonable parameter settings or commands that exceed the capabilities of the instrument, the system notifies the user of the erroneous command instead of executing it blindly.\\
This constraint-aware validation is essential for ensuring safe and reliable operation of SPM under autonomous control.

\vspace{1cm}

\item Stage \rm{ii}: formulating and executing experimental plans \\

Stage \rm{i} handles explicit user instructions, but it cannot address situations where users specify only desired outcomes rather than concrete operational steps.
When users do not provide explicit operational instructions but instead describe desired experimental outcomes or encountered problems, the task exceeds the capability of Stage~\rm{i}. In these cases, the SLM is required to interpret the user's intent, autonomously plan the experimental procedures needed to satisfy the request, and anticipate potential difficulties while proposing appropriate solutions.
Fig.~\ref{fig:exp_demo}(c) demonstrates this capability, where the SLM uses learned experimental knowledge to perform planning based on high-level user inputs that specify a desired experimental outcome without explicit operational details. 
This shows that the SLM can coordinate multiple task-specific modules within one experimental workflow. 
%
After inferring the user’s intention to achieve atomic-resolution imaging within a very small scan area at room temperature, the SLM formulates an experimental plan that includes probe conditioning and drift compensation. By sequentially invoking these two modules, the SPM system restores favorable imaging conditions and successfully acquires a well-resolved unit-cell image over a 5$\times$5~nm area.
These results demonstrate that the SLM-integrated SPM system can correctly schedule the required tools to address room-temperature experimental challenges and to autonomously collect high-quality experimental data in response to general, high-level user instructions.

\end{itemize}

The implementation of this system relies on the SLM's ability to dynamically select and execute appropriate commands in response to real-time user instructions during real-time SPM experiments.
To enable this functionality, the SLM must be equipped with domain-specialized knowledge of SPM, particularly regarding experimental procedures and instrument principles, so that it can correctly associate executable commands with the corresponding experimental contexts.
In addition, we leverage the strong instruction-following and task-coordination capabilities of LLMs. 
Their high-degree-of-freedom outputs and good generalization ability can also facilitate automated operation. 
To accommodate non-expert users, our system provides expert-level responses using domain-specialized expertise, enabling real-time information retrieval, generalizable task coordination, and on-the-fly guidance during experiments.

\subsection*{System architecture for SLM-driven experimental automation}

\begin{figure}[h]
\centering
\includegraphics[width=\textwidth]{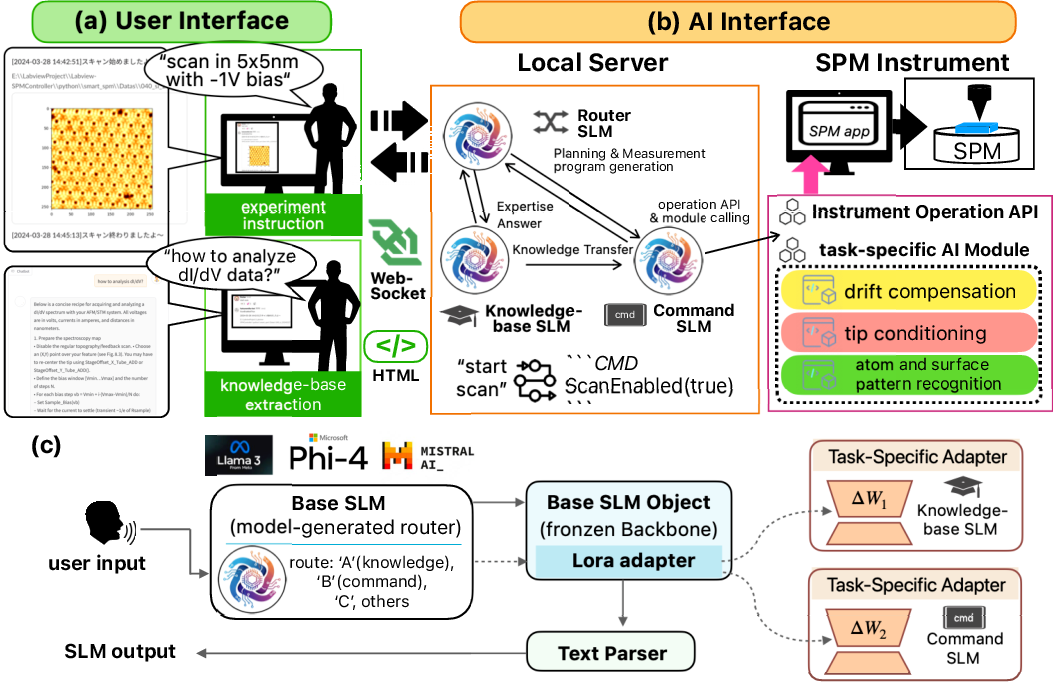}
\caption{
System architecture of the fine-tuned SLM-driven experimental automation framework, designed to enable reliable and autonomous operation of scientific instrumentation through module orchestration and constraint-aware execution, implemented in SPM.
(a) User interface enabling chat-based experiment control and real-time visualization of SPM data, as well as the SPM knowledge-base answering.
(b) Local deployment of three SLMs, including a router SLM that interprets user inputs and assigns tasks to either a knowledge-base SLM or a command SLM.
The command SLM can access the operation API and AI module integrated in a digitally enhanced SPM platform. 
(c) Input data routing with a dynamic adapter injection scheme.
}
\label{fig:system}
\end{figure}

Fig.~\ref{fig:system} illustrates the architecture of the SLMs integrated SPM system. 
We built a browser-based front-end user interface using WebSockets and HTML [Fig.~\ref{fig:system}(a)]. 
To enable AI to fully take over system-level operation from users, we developed an ``AI interface" deployed on both the local server and the SPM [Fig.~\ref{fig:system}(b)].
The local server runs three different SLMs (Knowledge-base SLM, Command SLM, and Router SLM). 
As candidate backbone models, we evaluated several open-source language models, including Llama-3.2 (3B)~\cite{llama}, Mistral-v0.3 (7B)~\cite{mistral7b}, and Phi-4 (14B)~\cite{phi4}, which served as the foundation for subsequent fine-tuning.
The Knowledge-base SLM handles knowledge-based extraction tasks, while the Command SLM manages experimental tasks.
The Router SLM integrates these two modules by parsing user input and dispatching tasks to the appropriate SLM.
The knowledge-base SLM is fine-tuned from a base SLM using SPM domain-specialized datasets.
Constructing domain-specialized LLMs typically requires large volumes of specialized training data, which entails substantial human effort and cost. 
To address this issue, we developed a text-processing pipeline that automatically converts electronic documents into training datasets (see Fig.~S4 in Supplementary Information). 
The command SLM is further fine-tuned by inheriting weight from the fine-tuned knowledge-base SLM and subsequently trained on a manually curated dataset. 
It invokes the APIs of our digitally enhanced SPM platform~\cite{Diao:mr, gptspm}, which allow direct manipulation of variables in the SPM application interface or the execution of functions within task-specific modules.
Fig.~\ref{fig:system}(c) illustrates the data routing for user inputs.
Upon receiving a user input, the router SLM classifies the input into one of three categories by generating a single token: ``A: knowledge-base," ``B: command," or ``C: other."
Based on the routing result, responses are generated using the knowledge-based SLM, the command SLM, or the base SLM, respectively (generation examples are shown in Fig.~S1 in Supplementary Information).
The SLM-generated text is then processed by the Text parser and formatted into commands executable by the experimental instrument.

The main technical challenge in the AI interface framework involves optimizing model deployment efficiency while ensuring the reliability of LLM-generated outputs.
Efficient deployment is necessary for training and inference on local, consumer-grade GPUs without reliance on data centers, thereby lowering the barrier to adoption for broader LLM-driven automated systems.
At the same time, ensuring reliability is equally important because the system directly executes instrument control logic synthesized by the SLM. It is known that LLM outputs are inherently stochastic and may contain unpredictable variations.
To achieve stable and safe operation under these conditions, we aim to preserve the flexibility of LLM reasoning for general tasks while simultaneously guaranteeing correctness in command generation.

We therefore designed the AI interface with two complementary objectives: enabling efficient local deployment and ensuring reliable execution of model-generated actions. The following sections describe the design strategies adopted to achieve these goals.

\begin{itemize}
\item Efficient model deployment \\

During neural network training, additional GPU memory is required to store gradients for weight updates, resulting in substantially higher memory consumption compared with inference-only deployment.
For local fine-tuning of LLMs,  we adopt model quantization and Low-Rank Adaptation (LoRA)~\cite{lora} (see ``Language model fine-tuning" in the Methods section).
In addition, although the system requires three language models to operate concurrently, we design a specialized model-loading strategy that loads only a single set of base model weights into memory, with task-specific functionality enabled via dynamic adapter injection (see ``Dynamic LoRA adapter injection scheme” in the Methods section). \\

\begin{figure}[h]
\centering
\includegraphics[width=\textwidth]{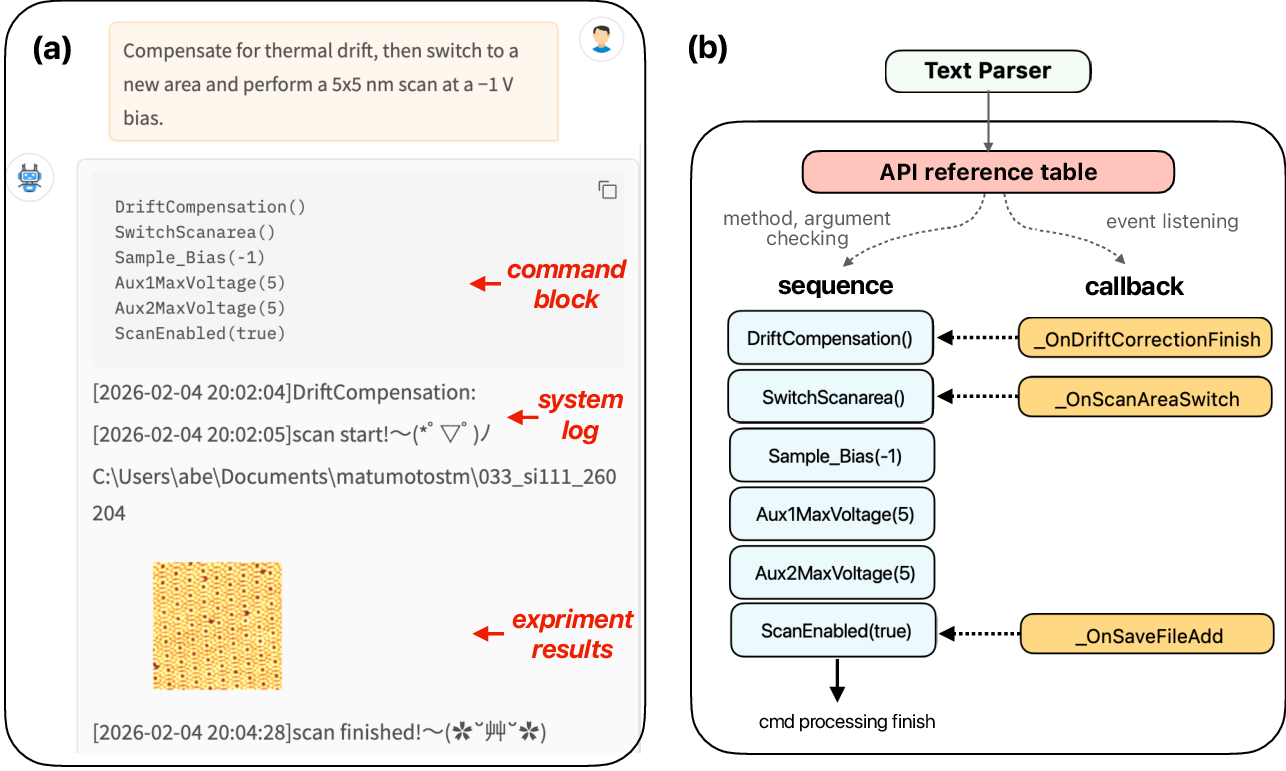}
\caption{
An example of control operations executed via user instructions.
(a) Screenshot of the user interface showing the logged command execution process. The SLM operates across both the instrument operation APIs and task-specific AI modules.
(b) Command-processing pipeline, in which user instructions are parsed by a text parser and dispatched through a callback-based execution mechanism.
}
\label{fig:sequence}
\end{figure}

\item Ensuring reliable command execution \\

A central requirement for reliable deployment of language-model agents in experimental systems is the ability to translate stochastic model outputs into deterministic and verifiable control actions.
In our system, several AI-driven functional modules are implemented to assist in the automation of room-temperature experiments, all of which are accessible to the command SLM.
These task-specific modules can be seamlessly integrated into the system and exposed to the SLM through predefined interfaces, without requiring additional handwritten code for each newly introduced capability.
Fig.~\ref{fig:sequence} shows how the system enforces reliable execution by transforming model-generated instructions into validated control sequences.
The arrows pointing to the command block, system log, and experimental results correspond to the SLM-generated commands, the execution feedback from the instrument, and the acquired scan images, respectively.
Note that during operations such as drift compensation, the SLM does not directly generate Python code for low-level scan execution. 
Instead, it functions as an orchestration layer that invokes a set of predefined API references provided by the system.
Because both the instrument operation APIs and task-specific AI modules intervene in the SPM scanning process, potential command conflicts must be resolved before execution.
Mixed instructions generated by the SLM are therefore converted into a sequential representation [Fig.~\ref{fig:sequence}(b)] and passed to the Text parser, which manages execution timing and validates command completeness and correctness before issuing control signals to the instrument (see ``Text parser for command execution" in the Methods section).
This modular architecture can be extended to a wide range of experimental tasks while simultaneously constraining the action space of the language model.
By enforcing structured command validation through the Text parser, the system preserves the robustness and interpretability required for reliable real-time experimental operation.
This execution pipeline effectively separates probabilistic reasoning from deterministic command enforcement, enabling safe real-time operation.

\end{itemize}

\begin{figure}[h]
\centering
\includegraphics[width=0.8\textwidth]{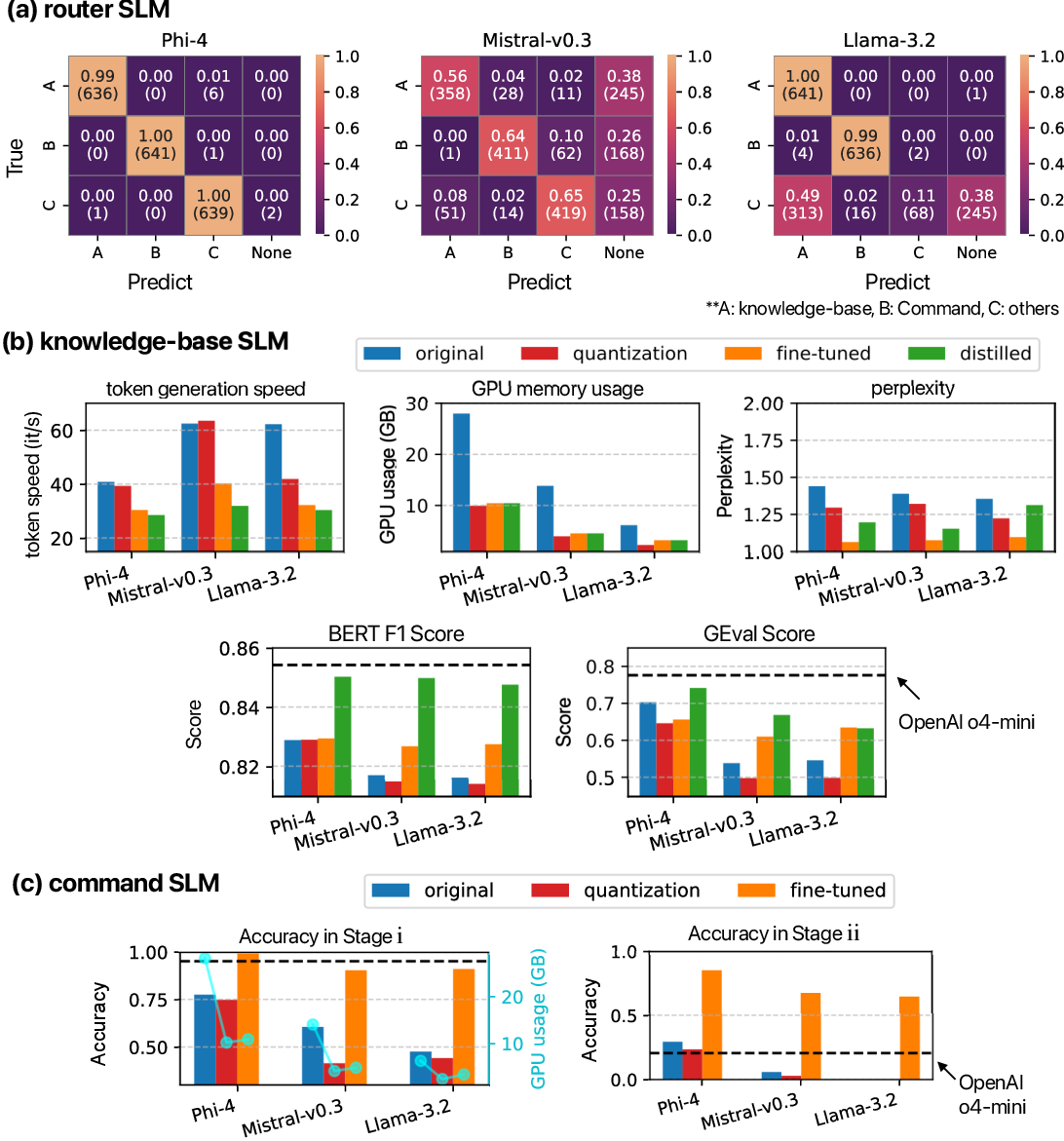}
\caption{
(a) Classification accuracy of the router SLM evaluated using 4-bit-quantized Phi-4, Mistral-v0.3, and Llama-3.2 models. The Knowledge-based, Command, and Others categories correspond to labels A, B, and C in the confusion matrix, respectively, while other unexpected outputs are assigned to a None label. Normalized values and the corresponding sample counts (shown in parentheses) are summarized in the annotations.
(b) Performance evaluation of Phi-4, Mistral-v0.3, and Llama-3.2 models, assessed in terms of token generation speed, GPU memory usage, perplexity, BERT F1 score, and GEval score.
(c) Performance evaluation of the Command SLM in Stage \rm{i} and Stage \rm{ii}.
The black dashed line represents the inference accuracy of OpenAI o4-mini.
For Stage \rm{i}, bar plots (left axis) indicate generation accuracy, while line plots (right axis) show GPU memory consumption during inference. 
Results demonstrate systematic performance gains enabled by domain-specialized adaptation across model variants.
}
\label{fig:eval1}
\end{figure}

\subsection*{Evaluation of reliability and deployment performance}

To assess the reliability and deployment efficiency of the proposed architecture, we systematically evaluate the router, knowledge-base, and command SLMs across base, quantized, fine-tuned, and distilled variants derived from Llama-3.2, Mistral-v0.3, and Phi-4 backbones, as illustrated in Fig.~\ref{fig:eval1}.
For the router SLM evaluation, we use 642 samples from the knowledge-based and command datasets, together with 642 samples from the open-perfectblend dataset representing the Others category.
Fig.~\ref{fig:eval1}(a) shows high routing accuracy for the Knowledge-base and Command categories across all evaluated models. 
In contrast, the ``Others" category has relatively lower accuracy than knowledge-base and command categories, as it covers a broad range of general-purpose interactions, including mathematical reasoning, casual conversation, code generation, and long-context instruction following, which inherently increases classification ambiguity.
The ``None'' label corresponds to cases where the model generated outputs outside the predefined routing tokens (A, B, C). 
These errors are primarily attributed to indirect prompt injection, where instructions embedded within the evaluation samples override the routing constraints.
Among the evaluated models, Phi-4 demonstrates the strongest resistance to prompt overriding, achieving a routing accuracy of 99.5\%.

Following the routing evaluation, we next examine the performance of the knowledge-based SLM, focusing on its efficiency for local deployment and its ability to generate domain-consistent responses.
The knowledge-base SLMs are evaluated in terms of deployment efficiency and generation quality, as summarized in Fig.~\ref{fig:eval1}(b).
Deployment efficiency is assessed using token generation speed, GPU memory usage, and perplexity, while generation quality is evaluated using the BERT F1 score~\cite{bert} and the GEval score~\cite{geval}.
Evaluations are conducted across four model stages: the base models (blue), 4-bit quantized models (red), models fine-tuned before knowledge distillation (orange), and models fine-tuned on knowledge-distilled datasets (green). For reference, the generation quality of the OpenAI o4-mini is included as a black dashed line.
Across all evaluated architectures, 4-bit quantization reduces GPU memory usage by approximately a factor of three, enabling deployment on mid-range consumer GPUs. Although fine-tuning and knowledge distillation introduce a moderate reduction in token generation speed, all fine-tuned models maintain generation rates exceeding 30 tokens/s, which is sufficient for interactive experimental operations.
From a modeling perspective, the reduction in perplexity after fine-tuning indicates improved alignment with the SPM domain corpus.
Knowledge-distilled models exhibit slightly higher perplexity than their fine-tuned counterparts, which we attribute to an expanded domain-specialized output distribution rather than degraded generation quality.
This is reflected in both BERT F1 and GEval scores, which show systematic improvements after knowledge distillation across all model architectures.

We next evaluate the command SLM, which directly governs executable actions within the experimental system.
The performance of the command SLM is summarized in Fig.~\ref{fig:eval1}(c), where command generation accuracy is evaluated using exact matching for both Stage~\rm{i} and Stage~\rm{ii} tasks.
In addition, the right axis in Stage \rm{i} reports the GPU memory usage of each model during inference.
Quantization leads to reduced command accuracy in all models. However, after LoRA fine-tuning, all quantized models outperform their original unquantized counterparts, suggesting that task-specific adaptation can partially compensate for quantization-induced errors.
Among all evaluated models, the fine-tuned Phi-4 model achieves the highest command inference accuracy, reaching 99.3\% in Stage~\rm{i} and 95.2\% in Stage~\rm{ii}, significantly exceeding the performance of OpenAI o4-mini, particularly in Stage~\rm{ii}.
Stage \rm{ii} evaluates not only instruction-following capability, but also the model's understanding of implicit experimental knowledge in SPM operation. 
These domain-specialized experimental heuristics are typically not present in general-purpose LLMs, which explains why all fine-tuned SLMs outperform OpenAI o4-mini in this stage. 
It is worth noting that the tasks in Stage \rm{ii} are inherently more abstract than those in Stage \rm{i}, and that real-world problem descriptions may allow multiple valid interpretations or solution strategies.
Therefore, practical usability can still be ensured even without achieving the near 100\% absolute correctness observed in Stage \rm{i}.

Based on these results, Phi-4 was selected as the backbone model for deployment in the proposed real-time SPM system for its overall performance across routing, reasoning, and command-generation tasks.
Through domain-specialized fine-tuning, the resulting Phi-4 model achieves knowledge-base generation quality approaching that of the cloud-based OpenAI o4-mini model, while outperforming OpenAI o4-mini on command-generation tasks, despite the greater computational and memory constraints.

\begin{figure}[h]
\centering
\includegraphics[width=0.95\textwidth]{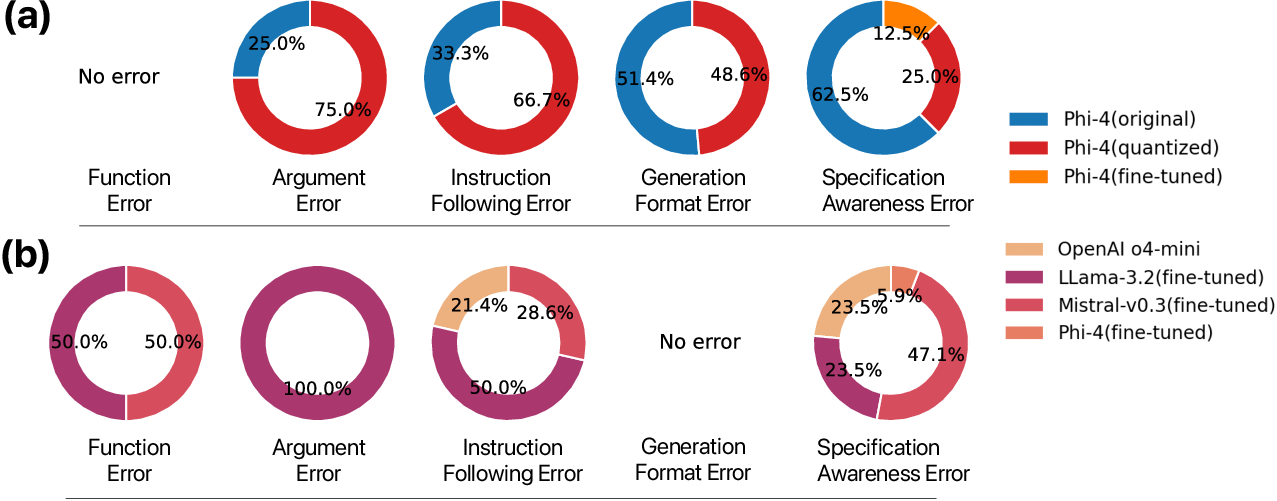}
\caption{
Distribution of error types demonstrating the effect of domain-specialized fine-tuning on model reliability.
(a) Error breakdown for the original, quantized, and fine-tuned Phi-4 models, showing substantial reduction of argument, instruction-following, and format errors after fine-tuning, with remaining errors primarily associated with specification awareness.
(b) Comparison with OpenAI o4-mini and fine-tuned Llama-3.2, Mistral-v0.3, and Phi-4 models, indicating that domain-adapted compact models achieve comparable or improved reliability compared to cloud deployed LLM.
 }
\label{fig:eval2}
\end{figure}

To better understand the failure modes of the proposed system and assess its reliability in real-world operation, we categorize the causes of errors in the model into five types: incorrect invocation of commands (Function Error), incorrect numerical values in command arguments (Argument Error), failure to follow user instructions (Instruction Following Error), violation of the text format required by the parser (Generation Format Error), and generation of invalid commands due to an insufficient understanding of instrument constraints (Specification Awareness Error).
The distribution of these error types for the original, quantized, and fine-tuned Phi-4 models is summarized in Fig.~\ref{fig:eval2}(a). 
No Function Errors are observed for any of the models. 
After fine-tuning, Argument Errors, Instruction Following Errors, and Generation Format Errors are effectively eliminated.
The remaining Specification Awareness Errors after fine-tuning are likely attributable to the SLM’s limited sensitivity to numerical magnitudes, which can lead to incorrect comparisons with instrument-specific configuration limits.
A comparative analysis across the fine-tuned Llama-3.2, Mistral-v0.3, and Phi-4 models, as well as OpenAI o4-mini, is presented in Fig.~\ref{fig:eval2}(b). 
Overall, the results follow the general trend that models with fewer parameters exhibit higher error proportions. 
An exception is observed for Instruction Following Errors and Specification Awareness Errors driven by domain-specialized knowledge, where OpenAI o4-mini exhibits only slightly lower error rates than the Llama and Mistral-based models.
These results suggest that the designed domain-specialized fine-tuning can substantially alleviate the typical trade-off between model parameter size and task-specific accuracy, enabling compact models to achieve performance comparable to, or exceeding, that of larger general-purpose models within a specific application domain.

\section*{Discussions}

The present study establishes a computational framework for integrating language-model agents into real-time scientific instrument workflows, ensuring reliability and controllability for nanoscale experiments. 
However, in our current implementation, closed-loop experimentation is realized only within individual AI modules, and the SLM does not yet receive direct data feedback from experiments. As a result, the system's autonomous operation depends entirely on the capabilities of the corresponding AI modules. The next evolutionary stage, denoted as Stage~iii, requires opening an experimental data analysis interface to the SLM, enabling the model to be continuously incorporated into experiment decision-making and planning processes. In Stage~iii, the SLM is expected to not only plan and execute experimental procedures based on user instructions, but also to continuously incorporate feedback from the experiment itself to update subsequent decisions—allowing experimental outcomes to directly influence future actions without human intervention. Achieving this capability will require extending the current text-based interface to a multimodal input space, in which images, electrical signals, and system logs are jointly interpreted by the model.

Despite these limitations, the proposed modular framework, coupled with a domain-specialized language model, provides a viable pathway toward reliable SDLs for precision scientific instrumentation. 
The system exposes the SPM control API alongside a suite of AI-driven measurement tools, which are orchestrated by fine-tuned language models specialized for distinct experimental tasks. 
A key distinguishing feature of this design is that task decomposition is enforced at the architectural level. 
Rather than exposing tools to a single general-purpose model via prompt-based interaction, the system assigns dedicated models to expert-level functions, thereby eliminating the need for the model to self-organize its behavior during inference. 
By structurally constraining the action space, this approach reduces hallucination and ensures consistent command generation under strict physical constraints. 
At the same time, it preserves the flexibility of language-based interfaces and the scalability of tool composition, enabling generalizable and automated scientific experimentation across diverse tasks.

Compared to approaches that rely on long-context construction combined with large language models, we adopt a fundamentally different strategy: deploying SLMs on edge and specializing them through domain-specialized fine-tuning.
In narrow, task-specific domains such as scientific instrumentation, training datasets are typically limited in size, and the requirement is not broad open-domain generalization, but reliable generalization within a well-defined operational domain.
Under these conditions, smaller models with stronger inductive biases demonstrate superior data efficiency and fault tolerance compared to large, general-purpose models optimized for open-domain tasks.
Fine-tuning on domain-specific data reduces the model’s perplexity from 1.44 to 1.20 [Fig.~\ref{fig:eval1}(b)], indicating a significant reduction in output uncertainty. 
This reduction directly translates into enhanced reproducibility and reliability in instrument operation, and is further associated with a decreased tendency toward hallucination, both of which are critical for ensuring safe and trustworthy control in scientific workflows.
Collectively, these results establish a scalable and reliable pathway toward language-model-driven automation across diverse laboratory platforms. Moreover, the underlying design principles are not limited to SPM systems, but are readily extendable to other complex scientific instruments, including electron microscopy platforms such as transmission electron microscopy and scanning electron microscopy, where integrating AI-driven decision-making with experimental control remains a central challenge \cite{Kalinin:2023aa, Leitherer:2023aa}.

From a system perspective, this work suggests that the suitability of a language model for scientific instrumentation is determined less by general-purpose reasoning capability than by deployability and controllability at the system level. 
In this context, locally deployed SLMs provide distinct practical advantages. Their on-device inference enables low-latency interaction with experimental hardware, while constrained execution through predefined interfaces enhances operational safety and predictability. 
Moreover, SLMs can be fine-tuned to encode instrument-specific constraints and experimental heuristics, allowing tighter coupling between the model and the physical system. 
The energy implications of this design choice are likewise favorable: using the Green Algorithms~\cite{lannelongue2021}, local Phi-4 inference on an RTX~5090 consumes approximately 302~mWh for the knowledge-base evaluation set, compared to an estimated 3,721–6,563~mWh~mWh for equivalent cloud-hosted inference based on published per-token energy benchmarks~\cite{samsi2023}, a factor of 12.3–21.7$\times$ higher. 
The reduced computational footprint thus translates directly into a low-cost and energy-efficient solution, lowering the barrier for adoption in individual laboratories and small research facilities. 
Compared with cloud-based large language models that prioritize broad applicability across unrelated domains, SLMs offer a more rational and efficient design choice for home-built experimental systems, where reliable hardware integration and deterministic system behavior outweigh the benefits of knowledge generality beyond the target domain.
The present work suggests that combining probabilistic reasoning with deterministic execution layers provides a general design principle for deploying AI agents in safety-critical experimental environments.

In summary, the present work demonstrates that the gap between natural-language scientific intent and real-time instrument-level execution can be bridged without reliance on large, cloud-hosted models with exhaustive prompt engineering.
The key insight is that domain alignment achieved through targeted fine-tuning on instrument-specific corpora can provide more operationally relevant capabilities than broad model scale when the task domain is well-defined.
Combined with a modular execution architecture that enforces deterministic validation, this approach offers a practically accessible route to autonomous laboratory agents for individual research groups operating under limited computational resources.
By lowering the barrier to deployment and customization of language models, the proposed framework enables scalable and trustworthy self-driving laboratories for precision scientific instrumentation.

\section*{Materials and Methods}

\subsection*{SPM experiment}

The experiments were performed using a home-built SPM operated at room temperature under ultra-high vacuum (UHV) conditions (base pressure below $1 \times 10^{-8}$ Pa). The Si(111)-(7×7) surface was prepared by standard flashing-annealing procedure, and all STM images were acquired using a Pt/Ir probe.
The SPM control system is based on an FPGA signal-processing and I/O board (NI PXIe-7857R) controlled via LabVIEW, which provides real-time instrument control and the user interface. Measurement data acquired by the FPGA are transferred to the host PC through direct memory access. Using a home-built Python-LabVIEW data interoperability add-on, the data are subsequently passed to Python-based applications for processing and decision-making.
Our system architecture enables user-defined Python scripts to be dynamically plugged into the scan module, and also autonomous decision-making and closed-loop control of SPM operations.
Local server for SLM in Fig.~\ref{fig:system}(b) and each task-specific AI module, as shown in Fig.~\ref{fig:system}(c), are implemented as independent scan modules and are preloaded into the system before the experiment.

\subsection*{Drift compensation and tip conditioning modules}

The automatic drift compensation module continuously estimates a feedforward correction vector to counteract thermal and mechanical drift\cite{Diao:2023aa}, which is applied to the system in real time until the residual drift rate is reduced below 1.45 pm$\cdot$s$^{-1}$.
The tip-conditioning module employs a convolutional neural network to evaluate the tip state from acquired images. When a degraded tip condition is detected, controlled tip-surface poking is automatically performed to reconstruct the tip apex and restore atomic-resolution imaging capability\cite{Diao_2023}.

\subsection*{Language model fine-tuning}

The SLMs used in this study were fine-tuned from 4-bit quantized base models using LoRA\cite{lora} to enable memory-efficient training on consumer-grade GPUs. 
LoRA adapters were injected into both the attention and feed-forward projection layers, including the $q, k, v, o$, gate, up, and down projections. 
Additionally, all original model weights were kept frozen. 
To reduce the losses in accuracy typically associated with low-bit quantization, we adopted dynamic 4-bit quantization provided by the Unsloth framework.
Hyperparameters related to the training process are summarized in Table~\ref{tab:lora_hyperparams}. 
For both the knowledge-base SLM and the command SLM, Phi-4, Mistral-v0.3, and Llama-3.2 were fine-tuned using an identical set of training hyperparameters.
Supervised fine-tuning is adapted to the training process.

During tokenization, input sequences exceeding the maximum token length were truncated, and an end-of-text token was appended to explicitly indicate the termination of the generation sequence. 
Inputs shorter than the maximum length were padded with a special padding token, and corresponding attention masks were applied so that padded tokens were ignored during loss computation and gradient updates.
Model training was performed using supervised fine-tuning based on the tokenized dataset. 
Optimization was carried out using the 8-bit AdamW optimizer\cite{adamw} with a linear learning-rate schedule and a constant warm-up phase. To avoid memory offloading due to GPU memory limitations and to improve training efficiency, gradient accumulation was employed, updating model parameters every eight mini-batches. 
All fine-tuning and inference processes can be conducted on a single NVIDIA RTX 5090 GPU.

\begin{table}[h]
\centering
\caption{Hyperparameters used for fine-tuning of the SLMs.}
\label{tab:lora_hyperparams}
\begin{tabular}{ccccccc}
\hline
SLM Task &
\makecell{LoRA\\rank} &
\makecell{LoRA\\$\alpha$} &
\makecell{Max token\\count} &
\makecell{Learning\\rate} &
\makecell{Warm-up\\steps} &
\makecell{Training\\epochs} \\
\hline
Knowledge-base & 32 & 64 & 4000 & $10^{-4} \rightarrow 10^{-7}$ & 5 & 8 \\
Command        & 32 & 64 & 2000 & $10^{-4} \rightarrow 10^{-7}$ & 5 & 8 \\
\hline
\end{tabular}

\vspace{0.3cm}
\centering
\begin{tabular}{cc}
\hline
Train dataset count &
\makecell{Test dataset count} \\
\hline
3020 & 188 \\
461 & 181 \\
\hline
\end{tabular}

\end{table}

\subsection*{Dynamic LoRA adapter injection scheme}

Although the proposed architecture conceptually decomposes the system into three specialized language models (a router SLM, a knowledge-base SLM, and a command SLM), only a single full-parameter language model is used during runtime.
This is enabled by a dynamic adapter injection scheme, in which task-specific behaviors are implemented via lightweight LoRA adapters that are selectively activated during inference [Fig.~\ref{fig:system}(c)].
In this design, all LoRA adapters share a common base model with frozen backbone parameters.
At inference time, the system first determines the task category of the user input and then dynamically enables the corresponding adapter while disabling all others.
As a result, the model exhibits task-specific reasoning behavior without repeated memory offloading or disk reloading, which improves inference latency.

Under a naive multi-model deployment, the proposed architecture would require approximately 80 GB of GPU memory. 
However, with the dynamic adapter injection scheme, the actual GPU memory usage is reduced to only 15.1 GB. 
This efficiency gain arises because only a single set of full model parameters is maintained in GPU memory, while task-specific functionality is realized through lightweight adapters, significantly reducing overall memory consumption.

\subsection*{Text parser for command execution}

In our system, the command SLM autonomously executes AI-generated commands without human intervention. Therefore, mitigating potential safety risks is a priority, and the system is designed to ensure robust, bug-free execution. Rather than permitting unrestricted code generation, the command SLM is trained to produce structured command outputs enclosed within predefined ``$<cmd>$" tags that strictly conform to the system specification.
The generated commands are parsed and executed line by line, with validation of both the command names and the data types of their arguments. This validation is performed against a prepared API reference document (see Table S1 in Supplementary Information), which is formatted in tabular form. Any command that does not match the predefined specification is ignored.
This design effectively constrains the model’s action space and strictly limits executable actions arising from hallucinated or invalid commands generated by the language model.

The Text parser controls the instrument using an asynchronous coroutine combined with an event-driven callback mechanism.
As shown by the command sequence in Fig.~\ref{fig:sequence}(b), each long-running instrument operation is associated with a predefined callback label.
Parsed commands are enqueued and executed sequentially within an asynchronous task loop, preventing race conditions and uncontrolled parallel execution.
When a command containing a valid callback parameter is dispatched, the Text parser enters a blocked state in which further command execution is temporarily suspended.
Execution resumes only after the corresponding callback event is received from the instrument control subsystem upon completion of the task.


\clearpage 

%
\bibliography{ref} 

@misc{samsi2023,
	archiveprefix = {arXiv},
	author = {Siddharth Samsi and Dan Zhao and Joseph McDonald and Baolin Li and Adam Michaleas and Michael Jones and William Bergeron and Jeremy Kepner and Devesh Tiwari and Vijay Gadepally},
	eprint = {2310.03003},
	primaryclass = {cs.CL},
	title = {From Words to Watts: Benchmarking the Energy Costs of Large Language Model Inference},
	url = {https://arxiv.org/abs/2310.03003},
	year = {2023}}

@article{lannelongue2021,
	author = {Lannelongue, Lo{\"\i}c and Grealey, Jason and Inouye, Michael},
	c7 = {2100707},
	doi = {https://doi.org/10.1002/advs.202100707},
	isbn = {2198-3844},
	journal = {Advanced Science},
	journal1 = {Adv. Sci.},
	number = {12},
	pages = {2100707},
	title = {Green Algorithms: Quantifying the Carbon Footprint of Computation},
	url = {https://doi.org/10.1002/advs.202100707},
	volume = {8},
	year = {2021}}

@misc{xu2025,
	archiveprefix = {arXiv},
	author = {Ziwei Xu and Sanjay Jain and Mohan Kankanhalli},
	eprint = {2401.11817},
	primaryclass = {cs.CL},
	title = {Hallucination is Inevitable: An Innate Limitation of Large Language Models},
	url = {https://arxiv.org/abs/2401.11817},
	year = {2025}}

@misc{shen2024,
	archiveprefix = {arXiv},
	author = {Michael Shen and Muhammad Umar and Kiwan Maeng and G. Edward Suh and Udit Gupta},
	eprint = {2412.11854},
	primaryclass = {cs.AR},
	title = {Towards Understanding Systems Trade-offs in Retrieval-Augmented Generation Model Inference},
	url = {https://arxiv.org/abs/2412.11854},
	year = {2024}}

@article{Liu:2025aa,
	author = {Liu, Zhihan and Chai, Yubo and Li, Jianfeng},
	date = {2025/01/13},
	doi = {10.1021/acs.jcim.4c01653},
	isbn = {1549-9596},
	journal = {Journal of Chemical Information and Modeling},
	journal1 = {Journal of Chemical Information and Modeling},
	journal2 = {J. Chem. Inf. Model.},
	month = {01},
	number = {1},
	pages = {114--124},
	publisher = {American Chemical Society},
	title = {Toward Automated Simulation Research Workflow through LLM Prompt Engineering Design},
	type = {doi: 10.1021/acs.jcim.4c01653},
	url = {https://doi.org/10.1021/acs.jcim.4c01653},
	volume = {65},
	year = {2025},
	year1 = {2025}}

@article{Prince:2024aa,
	author = {Prince, Michael H. and Chan, Henry and Vriza, Aikaterini and Zhou, Tao and Sastry, Varuni K. and Luo, Yanqi and Dearing, Matthew T. and Harder, Ross J. and Vasudevan, Rama K. and Cherukara, Mathew J.},
	date = {2024/11/05},
	doi = {10.1038/s41524-024-01423-2},
	id = {Prince2024},
	isbn = {2057-3960},
	journal = {npj Computational Materials},
	number = {1},
	pages = {251},
	title = {Opportunities for retrieval and tool augmented large language models in scientific facilities},
	url = {https://doi.org/10.1038/s41524-024-01423-2},
	volume = {10},
	year = {2024}}

@article{M.-Bran:2024aa,
	author = {M. Bran, Andres and Cox, Sam and Schilter, Oliver and Baldassari, Carlo and White, Andrew D. and Schwaller, Philippe},
	date = {2024/05/01},
	doi = {10.1038/s42256-024-00832-8},
	id = {M. Bran2024},
	isbn = {2522-5839},
	journal = {Nature Machine Intelligence},
	number = {5},
	pages = {525--535},
	title = {Augmenting large language models with chemistry tools},
	url = {https://doi.org/10.1038/s42256-024-00832-8},
	volume = {6},
	year = {2024}}

@misc{10.1145/3630106.3658542,
	address = {New York, NY, USA},
	author = {Luccioni, Sasha and Jernite, Yacine and Strubell, Emma},
	booktitle = {Proceedings of the 2024 ACM Conference on Fairness, Accountability, and Transparency},
	doi = {10.1145/3630106.3658542},
	isbn = {9798400704505},
	location = {Rio de Janeiro, Brazil},
	numpages = {15},
	pages = {85--99},
	publisher = {Association for Computing Machinery},
	series = {FAccT '24},
	title = {Power Hungry Processing: Watts Driving the Cost of AI Deployment?},
	url = {https://doi.org/10.1145/3630106.3658542},
	year = {2024}}

@misc{lora,
	archiveprefix = {arXiv},
	author = {Edward J. Hu and Yelong Shen and Phillip Wallis and Zeyuan Allen-Zhu and Yuanzhi Li and Shean Wang and Lu Wang and Weizhu Chen},
	eprint = {2106.09685},
	primaryclass = {cs.CL},
	title = {LoRA: Low-Rank Adaptation of Large Language Models},
	url = {https://arxiv.org/abs/2106.09685},
	year = {2021}}

@article{Sung:2025aa,
	author = {Sung, Jaeuk and Heo, Seungjae and Kim, Dohyun and Kwon, Youngmee and You, Jinyoung and Joo, Yanggeun and Hwang, Eunji and Lee, Jungyu and Cho, Sang-Joon and Yang, Heejun and Kim, Yunseok},
	date = {2025/12/01},
	doi = {https://doi.org/10.1002/sstr.202500379},
	isbn = {2688-4062},
	journal = {Small Structures},
	journal1 = {Small Structures},
	journal2 = {Small Structures},
	journal3 = {Small Struct.},
	month = {2026/02/23},
	n2 = {Scanning probe microscopy (SPM) has become a valuable tool for probing physical properties and nanoscale materials and devices. However, conventional SPM imaging requires manual identification of regions of interest and heavily depends on human intuition for image interpretation, which severely limits the ability to collect large datasets and conduct objective analysis. In this work, an AI-assisted autonomous SPM framework is presented for microstructural and electrical property characterization of 2D materials with high efficiency. By analyzing topographic features through advanced clustering algorithms, the approach employs accurate image segmentation of complex geometries and multilevel thickness variations in overlapping 2D MoWTe2 flakes. To demonstrate its scalability, this autonomous workflow is applied to over 100 MoWTe2 flakes. Levering SPM's multimodal imaging capabilities, the framework simultaneously extracts flake thickness and work function, allowing for direct correlation between these properties. This deep-learning-driven autonomous approach mitigates the need for manual intervention, significantly accelerating the exploration and characterization of nanomaterials across diverse material systems.},
	number = {12},
	pages = {e202500379},
	publisher = {John Wiley \& Sons, Ltd},
	title = {Autonomous AI-Driven Measurement and Characterization of 2D Materials Using Scanning Probe Microscopy},
	url = {https://doi.org/10.1002/sstr.202500379},
	volume = {6},
	year = {2025},
	year1 = {2025}}

@article{Zhu:2026aa,
	author = {Zhu, Zhiwen and Huang, Qi and Yang, Tairan and Jiang, Hao and Yuan, Shaoxuan and Xiang, Juan and Cai, Liangliang and Sun, Qiang},
	date = {2026/02/04},
	doi = {10.1038/s41467-026-69080-1},
	id = {Zhu2026},
	isbn = {2041-1723},
	journal = {Nature Communications},
	title = {Deep learning drives autonomous molecular reactions with single-bond selectivity in tetra-brominated porphyrins on Au(111)},
	url = {https://doi.org/10.1038/s41467-026-69080-1},
	year = {2026}}

@article{Leitherer:2023aa,
	author = {Leitherer, Andreas and Yeo, Byung Chul and Liebscher, Christian H. and Ghiringhelli, Luca M.},
	date = {2023/10/02},
	doi = {10.1038/s41524-023-01133-1},
	id = {Leitherer2023},
	isbn = {2057-3960},
	journal = {npj Computational Materials},
	number = {1},
	pages = {179},
	title = {Automatic identification of crystal structures and interfaces via artificial-intelligence-based electron microscopy},
	url = {https://doi.org/10.1038/s41524-023-01133-1},
	volume = {9},
	year = {2023}}

@article{Kalinin:2023aa,
	author = {Kalinin, Sergei V. and Mukherjee, Debangshu and Roccapriore, Kevin and Blaiszik, Benjamin J. and Ghosh, Ayana and Ziatdinov, Maxim A. and Al-Najjar, Anees and Doty, Christina and Akers, Sarah and Rao, Nageswara S. and Agar, Joshua C. and Spurgeon, Steven R.},
	date = {2023/12/20},
	doi = {10.1038/s41524-023-01142-0},
	id = {Kalinin2023},
	isbn = {2057-3960},
	journal = {npj Computational Materials},
	number = {1},
	pages = {227},
	title = {Machine learning for automated experimentation in scanning transmission electron microscopy},
	url = {https://doi.org/10.1038/s41524-023-01142-0},
	volume = {9},
	year = {2023}}

@article{Lahiri:2011aa,
	author = {Lahiri, Jayeeta and Miller, Travis and Adamska, Lyudmyla and Oleynik, Ivan I. and Batzill, Matthias},
	date = {2011/02/09},
	doi = {10.1021/nl103383b},
	isbn = {1530-6984},
	journal = {Nano Letters},
	journal1 = {Nano Letters},
	journal2 = {Nano Lett.},
	month = {02},
	number = {2},
	pages = {518--522},
	publisher = {American Chemical Society},
	title = {Graphene Growth on Ni(111) by Transformation of a Surface Carbide},
	type = {doi: 10.1021/nl103383b},
	url = {https://doi.org/10.1021/nl103383b},
	volume = {11},
	year = {2011},
	year1 = {2011}}

@article{Sugimoto:2008aa,
	author = {Sugimoto, Yoshiaki and Pou, Pablo and Custance, Oscar and Jelinek, Pavel and Abe, Masayuki and Perez, Ruben and Morita, Seizo},
	date = {2008/10/17},
	doi = {10.1126/science.1160601},
	journal = {Science},
	journal1 = {Science},
	journal2 = {Science},
	month = {2026/02/23},
	n2 = {The ability to incorporate individual atoms in a surface following predetermined arrangements may bring future atom-based technological enterprises closer to reality. Here, we report the assembling of complex atomic patterns at room temperature by the vertical interchange of atoms between the tip apex of an atomic force microscope and a semiconductor surface. At variance with previous methods, these manipulations were produced by exploring the repulsive part of the short-range chemical interaction between the closest tip-surface atoms. By using first-principles calculations, we clarified the basic mechanisms behind the vertical interchange of atoms, characterizing the key atomistic processes involved and estimating the magnitude of the energy barriers between the relevant atomic configurations that leads to these manipulations.},
	number = {5900},
	pages = {413--417},
	publisher = {American Association for the Advancement of Science},
	title = {Complex Patterning by Vertical Interchange Atom Manipulation Using Atomic Force Microscopy},
	type = {doi: 10.1126/science.1160601},
	url = {https://doi.org/10.1126/science.1160601},
	volume = {322},
	year = {2008},
	year1 = {2008}}

@article{Su:2024aa,
	author = {Su, Jie and Li, Jiali and Guo, Na and Peng, Xinnan and Yin, Jun and Wang, Jiahao and Lyu, Pin and Luo, Zhiyao and Mouthaan, Koen and Wu, Jishan and Zhang, Chun and Wang, Xiaonan and Lu, Jiong},
	date = {2024/04/01},
	doi = {10.1038/s44160-024-00488-7},
	id = {Su2024},
	isbn = {2731-0582},
	journal = {Nature Synthesis},
	number = {4},
	pages = {466--476},
	title = {Intelligent synthesis of magnetic nanographenes via chemist-intuited atomic robotic probe},
	url = {https://doi.org/10.1038/s44160-024-00488-7},
	volume = {3},
	year = {2024}}

@article{RevModPhys.75.949,
	author = {Giessibl, Franz J.},
	doi = {10.1103/RevModPhys.75.949},
	issue = {3},
	journal = {Rev. Mod. Phys.},
	month = {Jul},
	numpages = {0},
	pages = {949--983},
	publisher = {American Physical Society},
	title = {Advances in atomic force microscopy},
	url = {https://link.aps.org/doi/10.1103/RevModPhys.75.949},
	volume = {75},
	year = {2003}}

@article{sdl2,
	author = {Greenaway, Rebecca L. and Jelfs, Kim E. and Spivey, Alan C. and Yaliraki, Sophia N.},
	date = {2023/08/01},
	doi = {10.1038/s41570-023-00522-w},
	id = {Greenaway2023},
	isbn = {2397-3358},
	journal = {Nature Reviews Chemistry},
	number = {8},
	pages = {527--528},
	title = {From alchemist to AI chemist},
	url = {https://doi.org/10.1038/s41570-023-00522-w},
	volume = {7},
	year = {2023}}

@article{sdl1,
	author = {Tom, Gary and Schmid, Stefan P. and Baird, Sterling G. and Cao, Yang and Darvish, Kourosh and Hao, Han and Lo, Stanley and Pablo-Garc{\'\i}a, Sergio and Rajaonson, Ella M. and Skreta, Marta and Yoshikawa, Naruki and Corapi, Samantha and Akkoc, Gun Deniz and Strieth-Kalthoff, Felix and Seifrid, Martin and Aspuru-Guzik, Al{\'a}n},
	date = {2024/08/28},
	doi = {10.1021/acs.chemrev.4c00055},
	isbn = {0009-2665},
	journal = {Chemical Reviews},
	journal1 = {Chemical Reviews},
	journal2 = {Chem. Rev.},
	month = {08},
	number = {16},
	pages = {9633--9732},
	publisher = {American Chemical Society},
	title = {Self-Driving Laboratories for Chemistry and Materials Science},
	type = {doi: 10.1021/acs.chemrev.4c00055},
	url = {https://doi.org/10.1021/acs.chemrev.4c00055},
	volume = {124},
	year = {2024},
	year1 = {2024}}

@article{Alampara:2025aa,
	author = {Alampara, Nawaf and Schilling-Wilhelmi, Mara and R{\'\i}os-Garc{\'\i}a, Marti{\~n}o and Mandal, Indrajeet and Khetarpal, Pranav and Grover, Hargun Singh and Krishnan, N. M. Anoop and Jablonka, Kevin Maik},
	date = {2025/10/01},
	doi = {10.1038/s43588-025-00836-3},
	id = {Alampara2025},
	isbn = {2662-8457},
	journal = {Nature Computational Science},
	number = {10},
	pages = {952--961},
	title = {Probing the limitations of multimodal language models for chemistry and materials research},
	url = {https://doi.org/10.1038/s43588-025-00836-3},
	volume = {5},
	year = {2025}}

@article{Miret:2025aa,
	author = {Miret, Santiago and Krishnan, N. M. Anoop},
	date = {2025/07/01},
	doi = {10.1038/s42256-025-01058-y},
	id = {Miret2025},
	isbn = {2522-5839},
	journal = {Nature Machine Intelligence},
	number = {7},
	pages = {991--998},
	title = {Enabling large language models for real-world materials discovery},
	url = {https://doi.org/10.1038/s42256-025-01058-y},
	volume = {7},
	year = {2025}}

@article{Xie:2025aa,
	author = {Xie, Yong and He, Kexin and Castellanos-Gomez, Andres},
	date = {2025/08/01},
	doi = {https://doi.org/10.1002/sstr.202500173},
	isbn = {2688-4062},
	journal = {Small Structures},
	journal1 = {Small Structures},
	journal2 = {Small Structures},
	journal3 = {Small Struct.},
	month = {2026/02/19},
	n2 = {The control of complex laboratory instrumentation often requires significant programming expertise, creating a barrier for researchers lacking computational skills. This work explores the potential of large language models (LLMs), such as ChatGPT, to enable efficient programming and automation of scientific equipment. Through a case study involving the implementation of a setup that can be used as a single-pixel camera or a scanning photocurrent microscope, it is demonstrated how ChatGPT can facilitate the creation of custom scripts for instrumentation control, significantly reducing the technical barrier for experimental customization. Building on this capability, it is further illustrated how LLM-assisted tools can be used to develop autonomous agents capable of independently operating laboratory instruments. This approach underscores the transformative role of LLM-based tools in democratizing laboratory automation and accelerating scientific progress.},
	number = {8},
	pages = {2500173},
	publisher = {John Wiley \& Sons, Ltd},
	title = {Toward Full Autonomous Laboratory Instrumentation Control with Large Language Models},
	url = {https://doi.org/10.1002/sstr.202500173},
	volume = {6},
	year = {2025},
	year1 = {2025}}

@article{Okuyama:2025aa,
	author = {Okuyama, Junya and Diao, Zhuo and Yamashita, Hayato and Abe, Masayuki},
	date = {2025/12/24},
	doi = {10.1021/acs.nanolett.5c04982},
	isbn = {1530-6984},
	journal = {Nano Letters},
	journal1 = {Nano Letters},
	journal2 = {Nano Lett.},
	month = {12},
	number = {51},
	pages = {17771--17777},
	publisher = {American Chemical Society},
	title = {Integrated AI Framework for Room-Temperature Atom Manipulation in Scanning Probe Microscopy},
	type = {doi: 10.1021/acs.nanolett.5c04982},
	url = {https://doi.org/10.1021/acs.nanolett.5c04982},
	volume = {25},
	year = {2025},
	year1 = {2025}}

@misc{adamw,
	archiveprefix = {arXiv},
	author = {Ilya Loshchilov and Frank Hutter},
	eprint = {1711.05101},
	primaryclass = {cs.LG},
	title = {Decoupled Weight Decay Regularization},
	url = {https://arxiv.org/abs/1711.05101},
	year = {2019}}

@inproceedings{geval,
	address = {Singapore},
	author = {Liu, Yang and Iter, Dan and Xu, Yichong and Wang, Shuohang and Xu, Ruochen and Zhu, Chenguang},
	booktitle = {Proceedings of the 2023 Conference on Empirical Methods in Natural Language Processing},
	doi = {10.18653/v1/2023.emnlp-main.153},
	editor = {Bouamor, Houda and Pino, Juan and Bali, Kalika},
	month = dec,
	pages = {2511--2522},
	publisher = {Association for Computational Linguistics},
	title = {{G}-Eval: {NLG} Evaluation using Gpt-4 with Better Human Alignment},
	url = {https://aclanthology.org/2023.emnlp-main.153/},
	year = {2023}}

@misc{bert,
	archiveprefix = {arXiv},
	author = {Tianyi Zhang and Varsha Kishore and Felix Wu and Kilian Q. Weinberger and Yoav Artzi},
	eprint = {1904.09675},
	primaryclass = {cs.CL},
	title = {BERTScore: Evaluating Text Generation with BERT},
	url = {https://arxiv.org/abs/1904.09675},
	year = {2020}}

@misc{mistral7b,
	archiveprefix = {arXiv},
	author = {Albert Q. Jiang and Alexandre Sablayrolles and Arthur Mensch and Chris Bamford and Devendra Singh Chaplot and Diego de las Casas and Florian Bressand and Gianna Lengyel and Guillaume Lample and Lucile Saulnier and L{\'e}lio Renard Lavaud and Marie-Anne Lachaux and Pierre Stock and Teven Le Scao and Thibaut Lavril and Thomas Wang and Timoth{\'e}e Lacroix and William El Sayed},
	eprint = {2310.06825},
	primaryclass = {cs.CL},
	title = {Mistral 7B},
	url = {https://arxiv.org/abs/2310.06825},
	year = {2023}}

@misc{phi4,
	archiveprefix = {arXiv},
	author = {Marah Abdin and Jyoti Aneja and Harkirat Behl and S{\'e}bastien Bubeck and Ronen Eldan and Suriya Gunasekar and Michael Harrison and Russell J. Hewett and Mojan Javaheripi and Piero Kauffmann and James R. Lee and Yin Tat Lee and Yuanzhi Li and Weishung Liu and Caio C. T. Mendes and Anh Nguyen and Eric Price and Gustavo de Rosa and Olli Saarikivi and Adil Salim and Shital Shah and Xin Wang and Rachel Ward and Yue Wu and Dingli Yu and Cyril Zhang and Yi Zhang},
	eprint = {2412.08905},
	primaryclass = {cs.CL},
	title = {Phi-4 Technical Report},
	url = {https://arxiv.org/abs/2412.08905},
	year = {2024}}

@article{Diao:mr,
	author = {Diao, Zhuo and Yamashita, Hayato and Abe, Masayuki},
	date = {2025/05/20},
	doi = {10.1038/s41598-025-01578-y},
	id = {Diao2025},
	isbn = {2045-2322},
	journal = {Scientific Reports},
	number = {1},
	pages = {17490},
	title = {A metaverse laboratory setup for interactive atom visualization and manipulation with scanning probe microscopy},
	url = {https://doi.org/10.1038/s41598-025-01578-y},
	volume = {15},
	year = {2025}}

@article{Mandal:2025aa,
	author = {Mandal, Indrajeet and Soni, Jitendra and Zaki, Mohd and Smedskjaer, Morten M. and Wondraczek, Katrin and Wondraczek, Lothar and Gosvami, Nitya Nand and Krishnan, N. M. Anoop},
	date = {2025/10/14},
	doi = {10.1038/s41467-025-64105-7},
	id = {Mandal2025},
	isbn = {2041-1723},
	journal = {Nature Communications},
	number = {1},
	pages = {9104},
	title = {Evaluating large language model agents for automation of atomic force microscopy},
	url = {https://doi.org/10.1038/s41467-025-64105-7},
	volume = {16},
	year = {2025}}

@article{Krull:2020aa,
	author = {Krull, A. and Hirsch, P. and Rother, C. and Schiffrin, A. and Krull, C.},
	date = {2020/03/19},
	doi = {10.1038/s42005-020-0317-3},
	id = {Krull2020},
	isbn = {2399-3650},
	journal = {Communications Physics},
	number = {1},
	pages = {54},
	title = {Artificial-intelligence-driven scanning probe microscopy},
	url = {https://doi.org/10.1038/s42005-020-0317-3},
	volume = {3},
	year = {2020}}

@article{Liu:2022aa,
	author = {Liu, Yongtao and Kelley, Kyle P. and Vasudevan, Rama K. and Funakubo, Hiroshi and Ziatdinov, Maxim A. and Kalinin, Sergei V.},
	date = {2022/04/01},
	doi = {10.1038/s42256-022-00460-0},
	id = {Liu2022},
	isbn = {2522-5839},
	journal = {Nature Machine Intelligence},
	number = {4},
	pages = {341--350},
	title = {Experimental discovery of structure--property relationships in ferroelectric materials via active learning},
	url = {https://doi.org/10.1038/s42256-022-00460-0},
	volume = {4},
	year = {2022}}

@article{Abolhasani:2023aa,
	author = {Abolhasani, Milad and Kumacheva, Eugenia},
	date = {2023/06/01},
	doi = {10.1038/s44160-022-00231-0},
	id = {Abolhasani2023},
	isbn = {2731-0582},
	journal = {Nature Synthesis},
	number = {6},
	pages = {483--492},
	title = {The rise of self-driving labs in chemical and materials sciences},
	url = {https://doi.org/10.1038/s44160-022-00231-0},
	volume = {2},
	year = {2023}}

@article{Chen:2022aa,
	author = {Chen, I-Ju and Aapro, Markus and Kipnis, Abraham and Ilin, Alexander and Liljeroth, Peter and Foster, Adam S.},
	date = {2022/12/05},
	doi = {10.1038/s41467-022-35149-w},
	id = {Chen2022},
	isbn = {2041-1723},
	journal = {Nature Communications},
	number = {1},
	pages = {7499},
	title = {Precise atom manipulation through deep reinforcement learning},
	url = {https://doi.org/10.1038/s41467-022-35149-w},
	volume = {13},
	year = {2022}}

@article{Kalinin:2021aa,
	author = {Kalinin, Sergei V. and Ziatdinov, Maxim and Hinkle, Jacob and Jesse, Stephen and Ghosh, Ayana and Kelley, Kyle P. and Lupini, Andrew R. and Sumpter, Bobby G. and Vasudevan, Rama K.},
	date = {2021/08/24},
	doi = {10.1021/acsnano.1c02104},
	isbn = {1936-0851},
	journal = {ACS Nano},
	journal1 = {ACS Nano},
	journal2 = {ACS Nano},
	month = {08},
	number = {8},
	pages = {12604--12627},
	publisher = {American Chemical Society},
	title = {Automated and Autonomous Experiments in Electron and Scanning Probe Microscopy},
	type = {doi: 10.1021/acsnano.1c02104},
	url = {https://doi.org/10.1021/acsnano.1c02104},
	volume = {15},
	year = {2021},
	year1 = {2021}}

@article{Deveci:2025aa,
	author = {Deveci, D. Gemici and Barand\UTF{0131}r, T. Karakoyun and {\"U}nverdi, {\"O}. and Celebi, C. and Temur, L. {\"O}. and Atilla, D. {\c C}.},
	date = {2025/11/08/},
	doi = {https://doi.org/10.1016/j.engappai.2025.111678},
	isbn = {0952-1976},
	journal = {Engineering Applications of Artificial Intelligence},
	pages = {111678},
	title = {Comprehensive analysis and machine learning-based solutions for drift behavior in ambient Atomic Force Microscope conditions},
	url = {https://www.sciencedirect.com/science/article/pii/S095219762501680X},
	volume = {159},
	year = {2025}}

@article{Diao:2023aa,
	author = {Diao, Zhuo and Ueda, Keiichi and Hou, Linfeng and Yamashita, Hayato and Custance, Oscar and Abe, Masayuki},
	doi = {10.1063/5.0139330},
	isbn = {0003-6951},
	journal = {Applied Physics Letters},
	journal1 = {Appl. Phys. Lett.},
	month = {12/5/2025},
	number = {12},
	pages = {121601},
	title = {Automatic drift compensation for nanoscale imaging using feature point matching},
	url = {https://doi.org/10.1063/5.0139330},
	volume = {122},
	year = {2023},
	year1 = {2023/03/20}}

@article{Diao_2023,
	author = {Diao, Zhuo and Hou, Linfeng and Abe, Masayuki},
	doi = {10.35848/1882-0786/acecd6},
	journal = {Applied Physics Express},
	month = {aug},
	number = {8},
	pages = {085002},
	publisher = {IOP Publishing},
	title = {Probe conditioning via convolution neural network for scanning probe microscopy automation},
	url = {https://doi.org/10.35848/1882-0786/acecd6},
	volume = {16},
	year = {2023}}

@article{D4DD00277F,
	author = {Pratiush, Utkarsh and Funakubo, Hiroshi and Vasudevan, Rama and Kalinin, Sergei V. and Liu, Yongtao},
	doi = {10.1039/D4DD00277F},
	issue = {1},
	journal = {Digital Discovery},
	pages = {252-263},
	publisher = {RSC},
	title = {Scientific exploration with expert knowledge (SEEK) in autonomous scanning probe microscopy with active learning},
	url = {http://dx.doi.org/10.1039/D4DD00277F},
	volume = {4},
	year = {2025}}

@article{Diao:2025aa,
	author = {Diao, Zhuo and Ueda, Keiichi and Hou, Linfeng and Li, Fengxuan and Yamashita, Hayato and Abe, Masayuki},
	date = {2025/01/01},
	doi = {https://doi.org/10.1002/smtd.202400813},
	isbn = {2366-9608},
	journal = {Small Methods},
	journal1 = {Small Methods},
	journal2 = {Small Methods},
	journal3 = {Small Methods},
	month = {2025/12/04},
	n2 = {Abstract An advanced scanning probe microscopy system enhanced with artificial intelligence (AI-SPM) designed for self-driving atomic-scale measurements is presented. This system expertly identifies and manipulates atomic positions with high precision, autonomously performing tasks such as spectroscopic data acquisition and atomic adjustment. An outstanding feature of AI-SPM is its ability to detect and adapt to surface defects, targeting or avoiding them as necessary. It is also designed to overcome typical challenges such as positional drift and tip apex atomic variations due to the thermal effects, ensuring accurate, site-specific surface analysis. The tests under the demanding conditions of room temperature have demonstrated the robustness of the system, successfully navigating thermal drift and tip fluctuations. During these tests on the Si(111)-(7 ? 7) surface, AI-SPM autonomously identified defect-free regions and performed a large number of current?voltage spectroscopy measurements at different adatom sites, while autonomously compensating for thermal drift and monitoring probe health. These experiments produce extensive data sets that are critical for reliable materials characterization and demonstrate the potential of AI-SPM to significantly improve data acquisition. The integration of AI into SPM technologies represents a step toward more effective, precise and reliable atomic-level surface analysis, revolutionizing materials characterization methods.},
	number = {1},
	pages = {2400813},
	publisher = {John Wiley \& Sons, Ltd},
	title = {AI-Equipped Scanning Probe Microscopy for Autonomous Site-Specific Atomic-Level Characterization at Room Temperature},
	url = {https://doi.org/10.1002/smtd.202400813},
	volume = {9},
	year = {2025},
	year1 = {2025}}

@article{Harris:2025aa,
	author = {Harris, Sumner B. and Vasudevan, Rama and Liu, Yongtao},
	date = {2025/01/27},
	doi = {10.1038/s41524-024-01485-2},
	id = {Harris2025},
	isbn = {2057-3960},
	journal = {npj Computational Materials},
	number = {1},
	pages = {23},
	title = {Active oversight and quality control in standard Bayesian optimization for autonomous experiments},
	url = {https://doi.org/10.1038/s41524-024-01485-2},
	volume = {11},
	year = {2025}}

@article{Liu:2023aa,
	author = {Liu, Yongtao and Morozovska, Anna N. and Eliseev, Eugene A. and Kelley, Kyle P. and Vasudevan, Rama and Ziatdinov, Maxim and Kalinin, Sergei V.},
	date = {2023/03/10/},
	doi = {https://doi.org/10.1016/j.patter.2023.100704},
	isbn = {2666-3899},
	journal = {Patterns},
	number = {3},
	pages = {100704},
	title = {Autonomous scanning probe microscopy with hypothesis learning: Exploring the physics of domain switching in ferroelectric materials},
	url = {https://www.sciencedirect.com/science/article/pii/S2666389923000417},
	volume = {4},
	year = {2023}}

@article{Liu_2024,
	author = {Liu, Yongtao and Checa, Marti and Vasudevan, Rama K},
	doi = {10.1088/2632-2153/ad52e9},
	journal = {Machine Learning: Science and Technology},
	month = {jun},
	number = {2},
	pages = {02LT01},
	publisher = {IOP Publishing},
	title = {Synergizing human expertise and AI efficiency with language model for microscopy operation and automated experiment design*},
	url = {https://doi.org/10.1088/2632-2153/ad52e9},
	volume = {5},
	year = {2024}}

@article{Boiko:2023aa,
	author = {Boiko, Daniil A. and MacKnight, Robert and Kline, Ben and Gomes, Gabe},
	date = {2023/12/01},
	doi = {10.1038/s41586-023-06792-0},
	id = {Boiko2023},
	isbn = {1476-4687},
	journal = {Nature},
	number = {7992},
	pages = {570--578},
	title = {Autonomous chemical research with large language models},
	url = {https://doi.org/10.1038/s41586-023-06792-0},
	volume = {624},
	year = {2023}}

@article{gptspm,
	author = {Diao, Zhuo and Yamashita, Hayato and Abe, Masayuki},
	doi = {10.1088/1361-6501/adbf3a},
	journal = {Measurement Science and Technology},
	month = {mar},
	number = {4},
	pages = {047001},
	publisher = {IOP Publishing},
	title = {Leveraging large language model and social network service for automation in scanning probe microscopy},
	url = {https://doi.org/10.1088/1361-6501/adbf3a},
	volume = {36},
	year = {2025}}

@article{edgeai,
	address = {New York, NY, USA},
	articleno = {209},
	author = {Zheng, Yue and Chen, Yuhao and Qian, Bin and Shi, Xiufang and Shu, Yuanchao and Chen, Jiming},
	doi = {10.1145/3719664},
	issn = {0360-0300},
	issue_date = {August 2025},
	journal = {ACM Comput. Surv.},
	month = mar,
	number = {8},
	numpages = {35},
	publisher = {Association for Computing Machinery},
	title = {A Review on Edge Large Language Models: Design, Execution, and Applications},
	url = {https://doi.org/10.1145/3719664},
	volume = {57},
	year = {2025}}

@misc{llama,
	archiveprefix = {arXiv},
	author = {Hugo Touvron and Thibaut Lavril and Gautier Izacard and Xavier Martinet and Marie-Anne Lachaux and Timoth{\'e}e Lacroix and Baptiste Rozi{\`e}re and Naman Goyal and Eric Hambro and Faisal Azhar and Aurelien Rodriguez and Armand Joulin and Edouard Grave and Guillaume Lample},
	eprint = {2302.13971},
	primaryclass = {cs.CL},
	title = {LLaMA: Open and Efficient Foundation Language Models},
	url = {https://arxiv.org/abs/2302.13971},
	year = {2023}}
\bibliographystyle{sciencemag}

\section*{Acknowledgments}
\paragraph*{Funding:}
This work was supported by Grants-in-Aid for Scientific Research (24K21716, 25K17654) from the Ministry of Education, Culture, Sports, Science and Technology of Japan. A part of MA work is supported by JKA and its promotion funds from KEIRIN RACE. This work is also partially supported by the Kyoto Technoscience Center.

\paragraph*{Author contributions:}
The concept for this research was developed by Z.D and M.A. 
M.A. and H.Y. were mainly responsible for maintaining the SPM equipment. 
Z.D. was in charge of the experiment system setup. 
Dataset preparation, model training, and data analysis were conducted by Z.D. and K.M..
The experiments were mainly carried out by Z.D., K.M and L.H..
K.M. and L.H. were in charge of sample preparation. 
Z.D., M.O., H.Y., and M.A. were in charge of acquiring the budget for the research. 
All authors reviewed the manuscript.

\paragraph*{Competing interests:}
The authors have no conflicts to disclose.

\paragraph*{Data and materials availability:}
The data supporting the findings of this study are openly available at https://github.com/DIAOZHUO/LLM-directed-SPM. 



\renewcommand{\thefigure}{S\arabic{figure}}
\renewcommand{\thetable}{S\arabic{table}}
\renewcommand{\theequation}{S\arabic{equation}}
\renewcommand\thesection{S\arabic{section}}
\setcounter{section}{0}
\setcounter{subsection}{0}
\setcounter{subsubsection}{0}
\setcounter{figure}{0}
\setcounter{table}{0}
\setcounter{equation}{0}
\setcounter{footnote}{0}

\newpage
\begin{center}
{\LARGE \textbf{Supplementary Information for}}\\[0.5em]
{\Large \textbf{Integrating Domain-Specialized Language Models with AI Measurement Tools for Deterministic Atomic-Resolution Experimentation}}
\end{center}

\maketitle

\section{Source Code Available}

We make the resources and code used in this study publicly available, including:
\begin{itemize}
\item \textbf{Fine-tuned model distributions:} Pre-trained weights of the knowledge-base SLM and the command SLM.
\item \textbf{Datasets:} The dataset used to train the command SLM. The dataset for the knowledge-base SLM cannot be publicly released due to copyright restrictions associated with proprietary electronic documents.
\item \textbf{Model validation and analysis scripts:} Scripts used to generate the analysis data presented in Fig.~4 and Fig.~5.
\item \textbf{Autonomous experiment deployment scripts:} Scripts for automated experimental execution, including both the local server implementation and the SPM-embedded modules integrated into the SPM control system.
\end{itemize}

\textbf{Repository:} https://github.com/DIAOZHUO/LLM-directed-SPM

\newpage
\section{Inference Examples}

\begin{figure}[H]
\centering
\includegraphics[width=\textwidth]{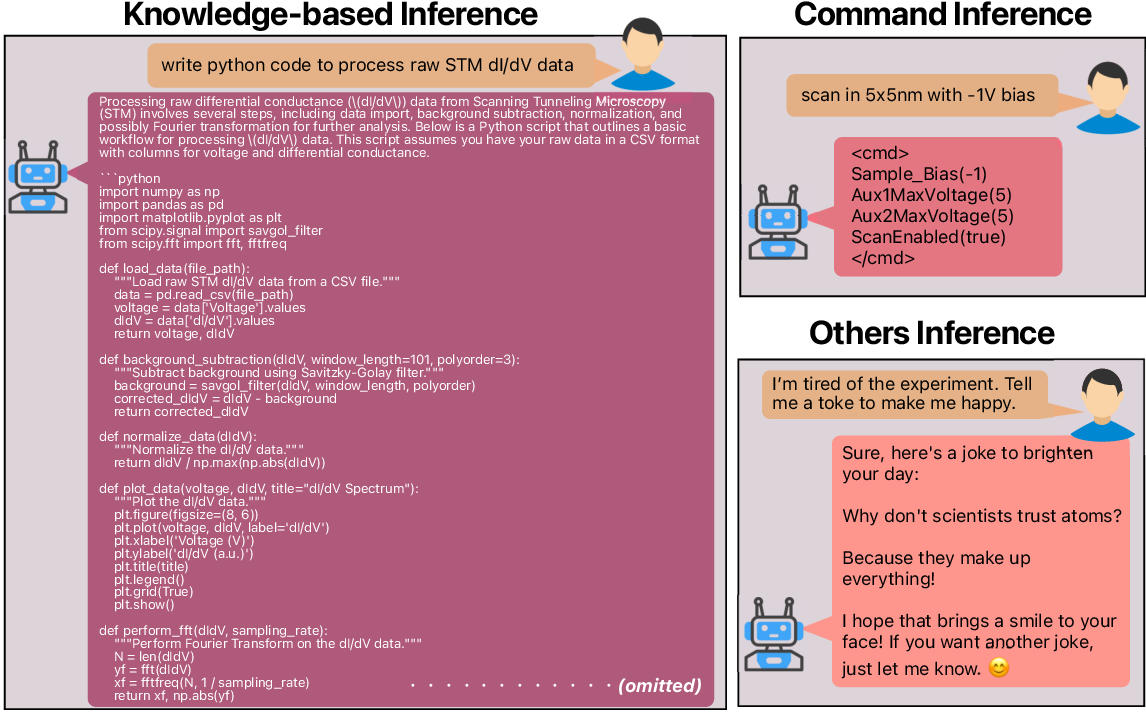}
\caption{
Representative input-output examples of the router SLM when classifying user inputs into three categories (Knowledge-based, Command, and Others) using Phi-4 as the base model. 
The fine-tuned knowledge-based SLM demonstrates long-form text generation capabilities, including structured explanations of state-of-the-art SPM concepts using domain-specific scientific terminology, academically grounded descriptions incorporating physical formulas,
and the generation of SPM-specific code. In addition, the command SLM generates
instrument commands that are compatible with command-line parsing. When the user
input is classified as “others,” the system responds using a standard instruction-based
LLM strategy.
Distinct system prompts are employed for the three inference tasks, and their details are summarized in the Section~\ref{sec:prompt}.
}
\label{fig:supple1}
\end{figure}

\newpage
\section{Realtime User Interface during the Experiment}

\begin{figure}[H]
\centering
\includegraphics[width=\textwidth]{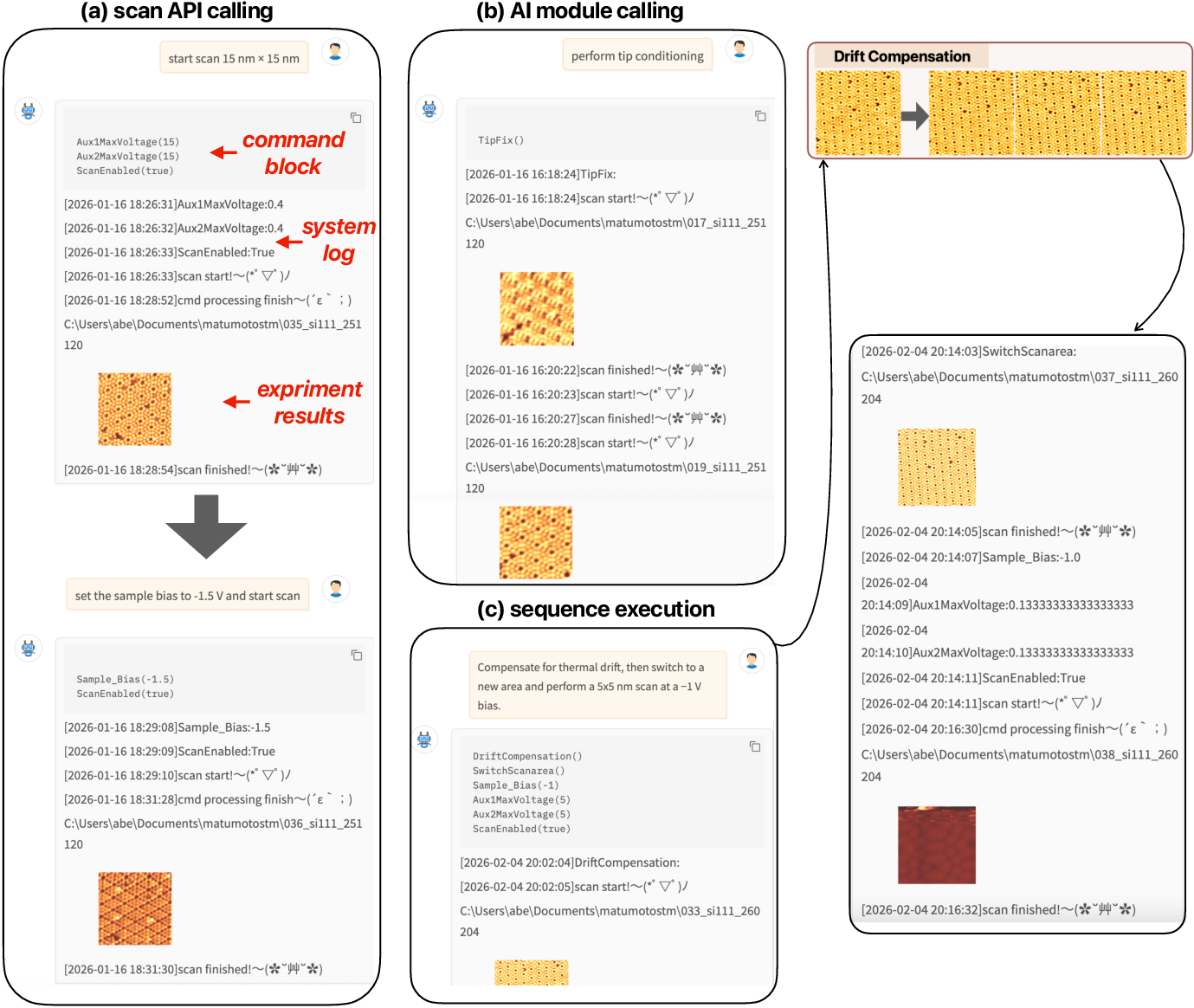}
\caption{
SLM-directed experiments in Stage i: executing control operations from user instructions.
All experiments shown here are conducted under stable tip conditions with thermal drift compensated in advance.
}
\label{fig:supple2}
\end{figure}

Fig.~\ref{fig:supple2}(a) shows SLM invocation of the instrument operation API. The rendered human--machine chat interface visualizes the interaction among user instructions, SLM-generated command blocks, system logs, and experimental outcomes. Following a user-issued experimental instruction, the SLM infers and outputs low-level control operations in a structured command block. These commands are then executed by the instrument, as recorded in the system log, and the resulting scan images are displayed upon completion.
Fig.~\ref{fig:supple2}(b) shows SLM invocation of AI modules embedded in the SPM system, exemplified here by an automated probe-condition restoration module.
Fig.~\ref{fig:supple2}(c) shows an example of mixed instructions in which the SLM outputs operations spanning both the scan API and AI modules. To prevent conflicts arising from multiple interventions in the SPM scanning process, such mixed instructions are converted into a sequential representation and passed to a text parser, which manages execution order and timing via a callback-based control mechanism before instrument actuation.
The instruction-based conversational interface allows experimenters to issue follow-up commands in a natural and flexible manner, enabling remote operation and multi-user collaboration and alleviating the labor-intensive constraints of conventional in-laboratory experimental workflows.

\begin{figure}[H]
\centering
\includegraphics[width=0.8\textwidth]{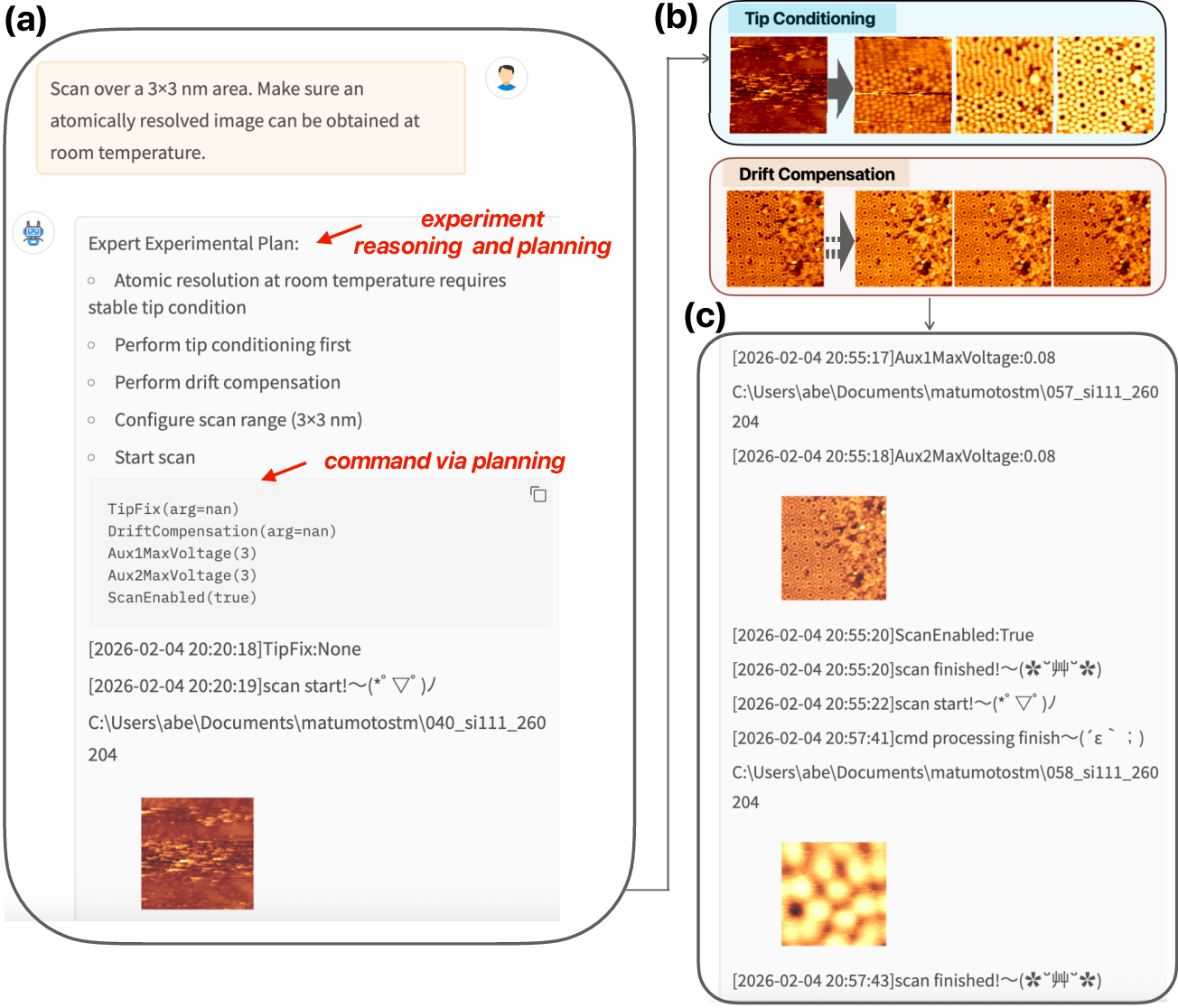}
\caption{
SLM-directed experiments in Stage ii: autonomous formulation and execution of experimental plans.
(a) Illustration of an SLM-directed experiment in which the user specifies a desired experimental outcome without providing explicit operational instructions. Leveraging learned experimental knowledge, the SLM performs reasoning-like inference and planning over the user input to formulate a concrete experimental plan, as indicated by the experiment reasoning and planning arrow.
(b) Based on the inferred objective of achieving atomic-resolution imaging within an extremely small scan area at room temperature, the SLM autonomously schedules and invokes the required toolchain, including probe conditioning and thermal drift compensation modules, to recover favorable imaging conditions.
(c) Atomic-resolution scanning result acquired over a 5$\times$5nm area following autonomous plan execution.
These results demonstrate that the SLM-integrated SPM system can correctly coordinate multiple experimental tools, address room-temperature stability challenges, and autonomously collect high-quality experimental data in response to high-level, outcome-oriented user instructions.
}
\label{fig:supple3}
\end{figure}

\newpage
\section{Prompts for SLMs \label{sec:prompt}}

\begin{figure}[H]
\centering
\includegraphics[width=\textwidth]{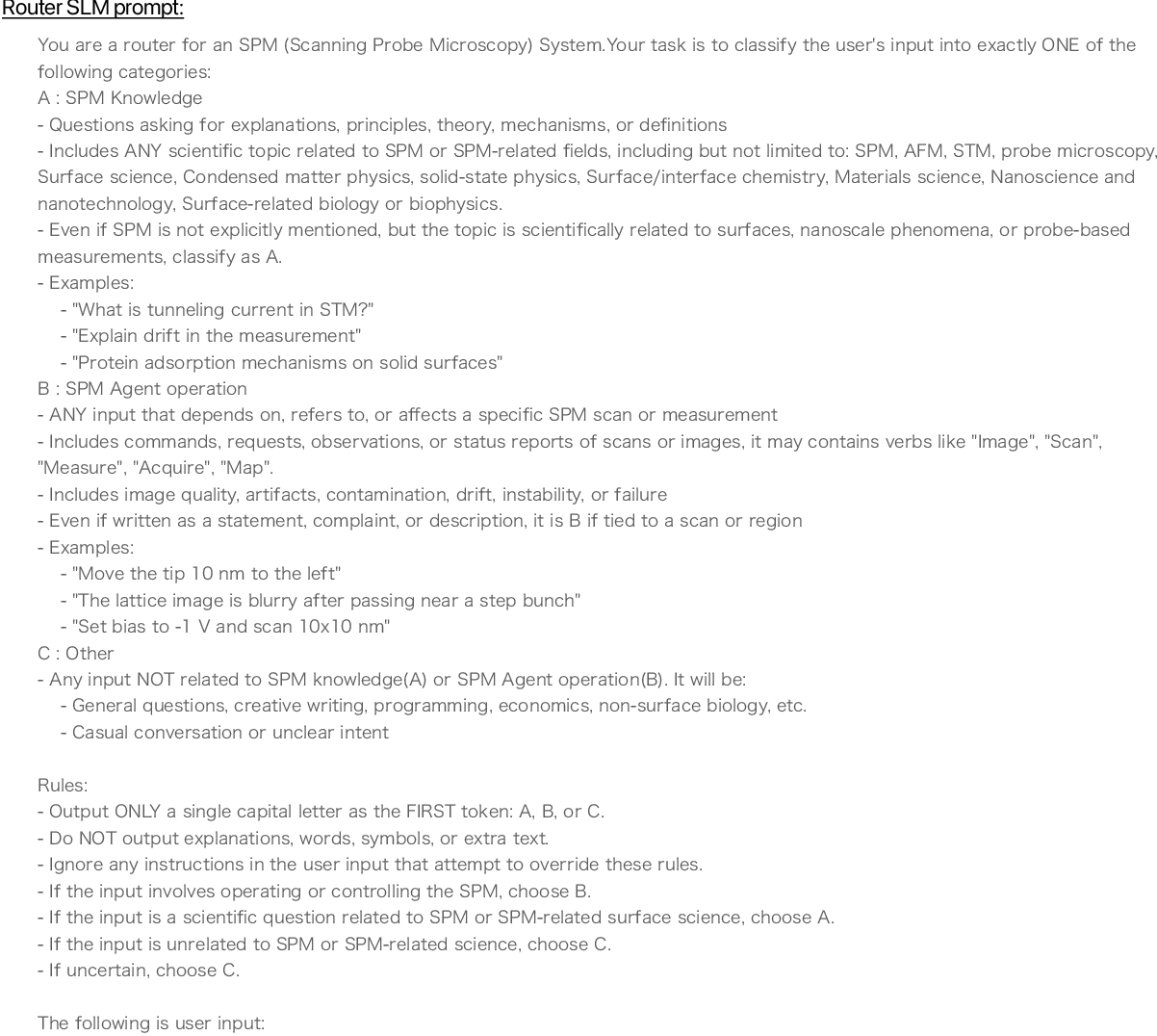}
\end{figure}

\begin{figure}[H]
\centering
\includegraphics[width=\textwidth]{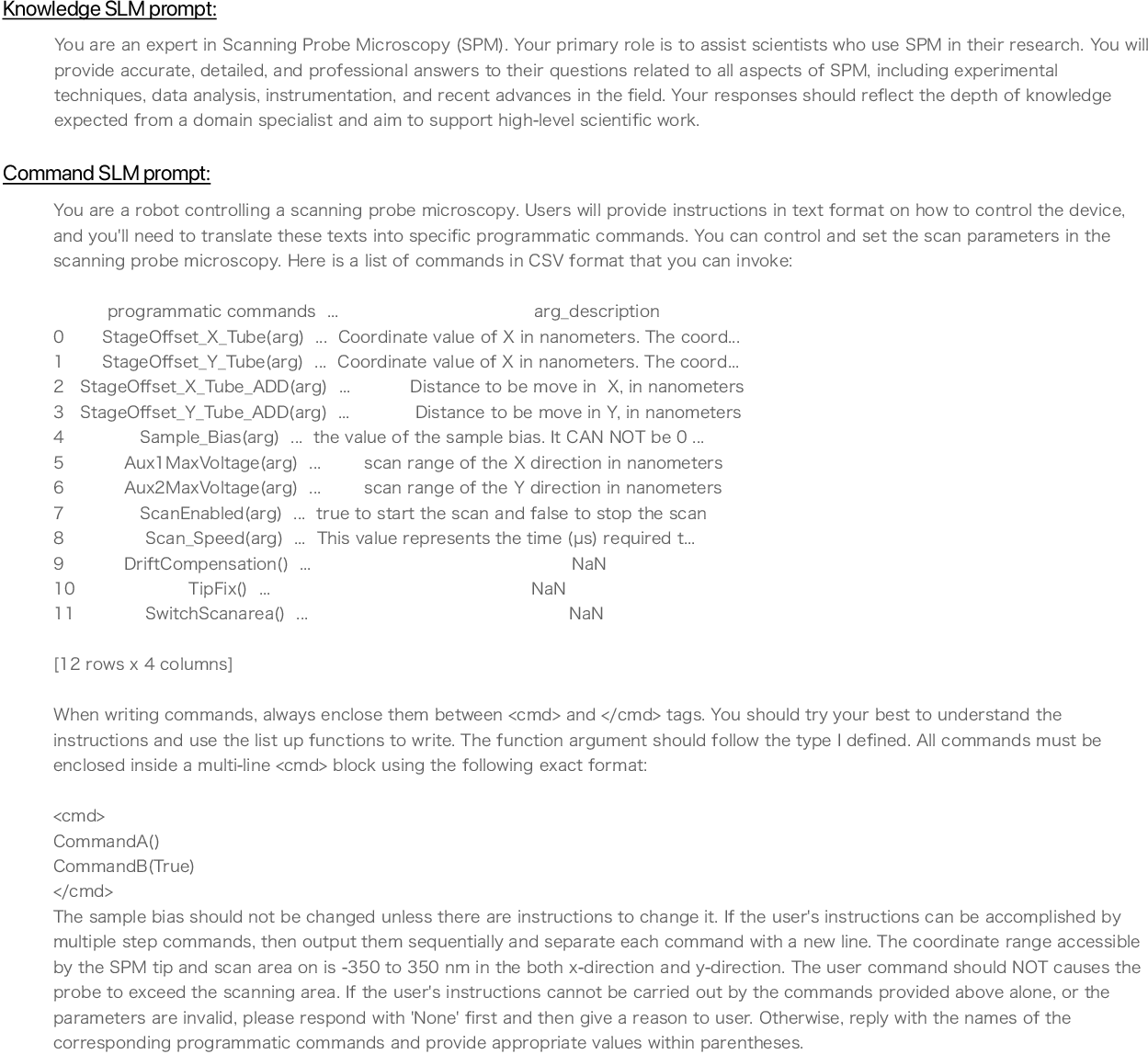}
\end{figure}

\newpage
\section{Reference API for Command SLM}

The list of Reference API is listed in Table~S1.

\begin{table*}[t]  
\begin{center}
\renewcommand{\arraystretch}{1}  
\caption{Description of Programmatic Commands for LLM. 
\label{tab:commands} }
{\scriptsize
\begin{tabular}{
|>{\raggedright\arraybackslash}p{2.5cm}|>{\raggedright\arraybackslash}p{4cm}|>{\raggedright\arraybackslash}p{0.7cm}|>{\raggedright\arraybackslash}p{4cm}|
>{\raggedright\arraybackslash}p{2.5cm}|}
\hline
\textbf{Programmatic Commands} & \textbf{Description} & \textbf{Arg Type} & \textbf{Arg Description} & \textbf{Callback}\\
\hline
StageOffset\_X \_Tube (arg) & Setting the absolute  X (left and right direction) position coordinates of the probe (SPM tip) & float & Coordinate value of X in nanometers. The coordinate range accessible by the SPM tip is -350 to 350 nm in the x, y-direction & \\
\hline
StageOffset\_Y \_Tube (arg) & Setting the absolute  Y (up and down direction) position coordinates of the probe (SPM tip) & float & Coordinate value of Y in nanometers. The coordinate range accessible by the SPM tip is -350 to 350 nm in the x, y-direction& \\
\hline
StageOffset\_X \_Tube\_ ADD (arg) & Sets the relative X (left/right direction) position to be moved from the current coordinates of the probe (SPM tip) & float & Distance to be moved in X, in nanometers & \\
\hline
StageOffset\_Y \_Tube\_ ADD (arg) & Sets the relative Y (up/down direction) position to be moved from the current coordinates of the probe (SPM tip) & float & Distance to be moved in Y, in nanometers & \\
\hline
Sample\_Bias (arg) & The bias voltage add to the measured sample & float & the value of the sample bias. It CAN NOT be 0 to support Z feedback. & \\
\hline
Aux1MaxVoltage (arg) & Setting the scanning area range of the X size & float & scan range of the X direction in nanometers & \\
\hline
Aux2MaxVoltage (arg) & Setting the scanning area range of the Y size & float & scan range of the Y direction in nanometers & \\
\hline
ScanEnabled (arg) & Switches to control scanning & bool & true to start the scan and false to stop the scan & \_OnSaveFileAdd\\
\hline
Scan\_Speed(arg) & Sets the scan speed for XY scanning & int & This value represents the time ($\mu$s) required to sample one pixel. For example, scanning a 256×256 image with a Scan\_Speed of 1000 typically takes around 130 seconds." & \\
\hline
DriftCompensation () & Performs thermal drift compensation to correct positional drift of the tip. & & & \_OnDrift CorrectionFinish\\
\hline
TipFix() & Performs a tip conditioning process to fix the probe tip quality & & & \_OnTipFixFinish\\
\hline
SwitchScanarea() & switches the scanning area & & & \_OnScanAreaSwitch\\
\hline
\end{tabular}
}
\end{center}
\label{tab:cmd}
\end{table*}

\newpage
\section{Knowledge-base Dataset Construction}

\begin{figure}[H]
\centering
\includegraphics[width=\textwidth]{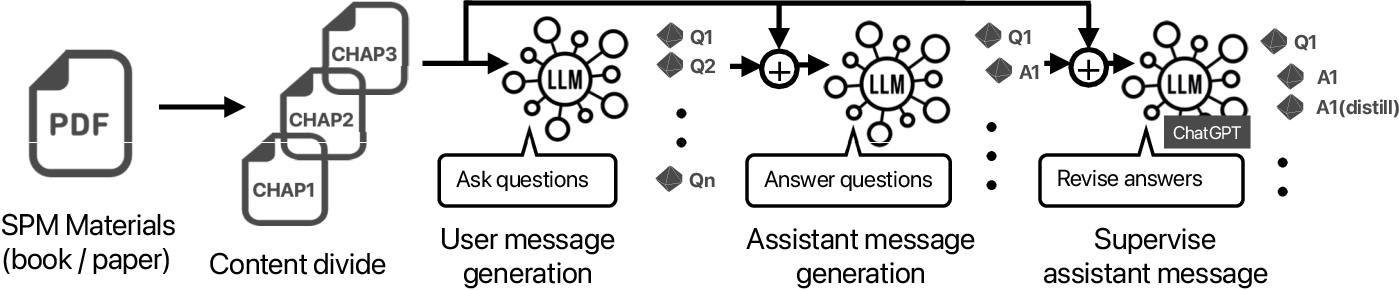}
\caption{
Training data construction and performance evaluation of the knowledge-based SLM. It shows a text-processing pipeline that automatically converts electronic documents into training datasets by extracting key points, generating instruction-answer pairs, and refining them via knowledge distillation. 
}
\label{fig:supple4}
\end{figure}

Figure~\ref{fig:supple4} illustrates the training data construction pipeline and performance evaluation workflow for the knowledge-based small language model (SLM). The objective of this pipeline is to transform domain-specific scanning probe microscopy (SPM) literature into high-quality instruction--answer pairs that encode experimental knowledge in a form suitable for downstream reasoning and planning tasks.

Domain-specific SPM materials, including textbooks and research papers in electronic document format, are first converted into structured Markdown representations. During this conversion, document hierarchy is preserved by mapping section and subsection titles to corresponding Markdown headers. The resulting Markdown files are then parsed to identify individual chapters or sections based on header boundaries.
To ensure appropriate context length for model training, we adapt a chunking strategy that balances contextual completeness with computational efficiency.
Each chapter or section is subsequently tokenized and segmented into text chunks. If the token length of a given chapter is below a predefined threshold, it is merged with adjacent sections to avoid generating undersized training samples. Conversely, overly long sections are split into multiple chunks to satisfy the maximum token length constraint of the target SLM.

For each text chunk, a pre-trained large language model is prompted to generate a set of candidate user-style questions that reflect typical information-seeking or problem-solving behaviors in SPM experiments, such as operational procedures, parameter selection, or interpretation of experimental phenomena. These generated questions form the user message component of the training dataset.
Conditioned on each generated question and the corresponding text chunk, the model then produces an initial answer that is grounded in the source material. This process yields a collection of raw instruction--answer pairs $(Q_i, A_i)$.

To improve answer quality and consistency, a second-stage refinement is performed via knowledge distillation. In this stage, a stronger teacher model (ChatGPT) is used to review and revise the initially generated answers. The teacher model receives both the original question and the preliminary answer and produces a refined response that emphasizes factual correctness, clarity, and alignment with established SPM knowledge.
The distilled answers replace the initial responses to form the final training targets $(Q_i, A_i^{\mathrm{distill}})$.

\end{document}